\documentclass[aps]{revtex4}

\usepackage{amsmath}

\def\ov#1{\overline{#1}}

\def\wt#1{\widetilde{#1}}
\def\vb#1{\mbox{\boldmath$#1$}}
\def\pd#1#2{\frac{\partial #1}{\partial #2}}
\def\fd#1#2{\frac{\delta #1}{\delta #2}}
\def\wh#1{\widehat{#1}}
\def\bdot{\,\vb{\cdot}\,}
\def\btimes{\,\vb{\times}\,}

\def\bhat{\wh{{\sf b}}}
\def\exd{{\sf d}}

\newcommand{\bc}{\begin{center}}
\newcommand{\ec}{\end{center}}
\newcommand{\bt}{\begin{tabbing}}
\newcommand{\et}{\end{tabbing}} 
\newcommand{\be}{\begin{eqnarray*}}
\newcommand{\ee}{\end{eqnarray*}}

\begin{document}

\title{Exact energy conservation laws for full and truncated nonlinear gyrokinetic equations}

\author{Alain J.~Brizard}
\affiliation{Department of Physics, Saint Michael's College, Colchester, VT 05439, USA \\
and CEA, IRFM, F-13108, Saint-Paul-lez-Durance, France}

\begin{abstract}
The exact global energy conservation laws for the full and truncated versions of the nonlinear electromagnetic gyrokinetic equations in general magnetic geometry are presented. In each version, the relation between polarization and magnetization effects in the gyrokinetic Poisson and Amp\`{e}re equations and the quadratic ponderomotive gyrocenter Hamiltonian is emphasized.
\end{abstract}

\begin{flushright}
March 2, 2010
\end{flushright}


\maketitle

\section{Introduction}

Nonlinear gyrokinetic Vlasov-Maxwell theory \cite{Brizard_Hahm} presents a powerful paradigm for the theoretical and numerical investigations of low-frequency turbulent phenomena in strongly magnetized plasmas. The gyrokinetic Maxwell-Vlasov equations describe the self-consistent coupling between the gyrocenter Vlasov distribution $F$ (in a reduced phase space where the gyrocenter magnetic moment is an invariant and the gyrocenter gyroangle is ignorable) and low-frequency electromagnetic fluctuations ${\bf E}_{1} \equiv -\,\nabla\phi_{1} - c^{-1}\partial{\bf A}_{1}/\partial t$ and ${\bf B}_{1} \equiv \nabla\btimes {\bf A}_{1}$ (which satisfy the gyrokinetic Maxwell's equations), in the presence of a nonuniform, time-independent background magnetic field ${\bf B}_{0}$. First, the background magnetic field ${\bf B}_{0} \equiv B_{0}\,\bhat_{0}$ is assumed to satisfy standard space-time-scale orderings based on guiding-center Hamiltonian theory \cite{RGL_83,Cary_Brizard}, which makes gyrokinetic theory versatile enough to be modified for special plasma geometries (e.g., edge plasmas \cite{Brizard_Hahm,edge_1,edge_2,HWM}). Next, according to the low-frequency gyrokinetic ordering \cite{Brizard_Hahm}, it is customary to use the following approximate expressions for the perturbed electric field: ${\bf E}_{1\bot} \simeq \nabla_{\bot}
\phi_{1}$ and $E_{1\|} = -\,\bhat_{0}\bdot\nabla\phi_{1} - c^{-1}\partial A_{1\|}/\partial t$. We also use the gyrokinetic fluctuation orderings
\cite{Brizard_Hahm} $c\,|{\bf E}_{1\bot}|/B_{0} \sim \epsilon\,v_{\rm th}$, $|E_{1\|}| \sim \epsilon\,|{\bf E}_{1\bot}|$, and $|{\bf B}_{1}| \sim \epsilon\,B_{0}$, where $\epsilon \ll 1$ is defined in terms of the time-scale ordering for the characteristic fluctuation frequency $\omega \sim \epsilon\,|\Omega|$ ($\Omega \equiv e\,B_{0}/mc$ is the gyrofrequency of a particle of mass $m$ and charge $e$ and $v_{\rm th}$ is the thermal velocity).

\subsection{Energy Conservation}

Gyrokinetic theory provides a reduced dynamical description of nonlinear (low-frequency) plasma turbulence that has important analytical and numerical advantages (see Ref.~\cite{Brizard_Hahm} for further details and references). The property of global energy conservation of the gyrokinetic equations, in particular, is often viewed as an important test for a nonlinear gyrokinetic numerical simulation code. While the total energy of the Vlasov-Maxwell equations in particle phase space is easily separated into the total kinetic energy of the particles and the electromagnetic field energy, the explicit decomposition of the total energy of reduced Vlasov-Maxwell equations can be far less obvious. This is because the treatment of the polarization and magnetization associated with the dynamical reduction of the Vlasov equation is, in general, complicated \cite{Brizard_JCPS}. Fortunately, the situation is not hopeless since various important sets of reduced Vlasov-Maxwell equations can be derived from reduced variational principles \cite{Brizard_VP1,Brizard_VP2,Sugama}. Hence, through the Noether method \cite{Brizard_05}, exact conservation laws can be derived for each set of reduced Vlasov-Maxwell equations, even though these equations are themselves approximate.

The nonlinear gyrokinetic Vlasov-Poisson \cite{Dubin,Hahm_88,Hahm_96} and Vlasov-Maxwell \cite{HLB,Brizard_89,HWM} equations can be numerically implemented by following either a ``full-$f$'' version \cite{fullf_1,fullf_2} or a truncated (``$\delta f$'') version \cite{deltaf_1,deltaf_2}. The purposes of the present work is (i) to show that each version is associated with a numerical implementation that possesses an exact gyrokinetic energy conservation law and (ii) to highlight the role played by the gyrocenter polarization and magnetization in ensuring exact energy conservation. 

\subsection{organization}

The remainder of the paper is organized as follows. In Sec.~\ref{sec:gyro_Ham}, we present a brief review of standard Hamiltonian gyrokinetic 
theory \cite{Brizard_Hahm}, where the gyrocenter Hamiltonian dynamics is expressed in terms of the unperturbed guiding-center Poisson bracket and the gyrocenter Hamiltonian (whose first and second-order terms represent the effects of the fluctuating fields). In Sec.~\ref{sec:gyro_polmag}, we introduce explicit expressions for the first-order gyrocenter polarization and magnetization associated with the gyrocenter transformation. In particular, we show how these terms appear in the second-order gyrocenter Hamiltonian defined in Sec.~\ref{sec:gyro_Ham}. In Sec.~\ref{sec:full}, we briefly review the 
variational derivation of the full-$f$ version of the nonlinear gyrokinetic equations and discuss the connection between the second-order ponderomotive term in the gyrocenter Hamiltonian and the property of exact energy conservation. First, we introduce the gyrocenter kinetic energy $K_{\rm gy} \equiv \langle{\sf T}_{\rm gy}^{-1}H_{0{\rm gc}}\rangle$, defined as the gyroangle-averaged part of the gyrocenter push-forward ${\sf T}_{\rm gy}^{-1}$ of the (unperturbed) guiding-center kinetic energy $H_{0{\rm gc}}$. Next, we present a new proof of exact energy conservation that emphasizes the role played by the second-order gyrocenter Hamiltonian. 

In Sec.~\ref{sec:trunc}, we present the variational derivation of the truncated version of the gyrokinetic Vlasov-Maxwell equations (in which the second-order gyrocenter Hamiltonian is omitted from the truncated gyrokinetic Vlasov equation), while the gyrokinetic Maxwell's equations retain the gyrocenter polarization and magnetization effects. From this truncated variational principle, we obtain an exact energy conservation law (by Noether method), which is also proved explicitly. Lastly, we summarize our work and discuss applications in Sec.~\ref{sec:summ}.

\section{\label{sec:gyro_Ham}Hamiltonian Gyrocenter Theory}

In this Section, we review the standard nonlinear Hamiltonian gyrokinetic theory, as derived by phase-space-Lagrangian Lie-transform perturbation method (additional details and references can be found in Ref.~\cite{Brizard_Hahm}). Before we begin with this review, we make a few remarks concerning the general theory of Hamiltonian dynamics in eight-dimensional extended phase space \cite{Brizard_Vlasovia}, where the coordinates $z^{a} \equiv ({\bf x},
{\bf p}; w, t)$ include the energy $w$ and time $t$ as canonically-conjugate coordinates.

The extended Hamilton's equations are derived from the variational principle $\delta\int\,(\Gamma - {\cal H}\,\exd\sigma) = 0$, where $\sigma$ denotes the Hamiltonian orbit parameter and the extended phase-space Lagrangian $\Gamma - {\cal H}\,\exd\sigma$ is defined in terms of the symplectic one-form 
$\Gamma \equiv \Gamma_{a}\,\exd z^{a}$ (summation over repeated indices is henceforth implied) and the extended Hamiltonian ${\cal H} \equiv H({\bf x},
{\bf p}, t) - w$. From this variational principle, we obtain the Euler-Lagrange equations $\omega_{ab}\,dz^{b}/d\sigma = \partial{\cal H}/\partial 
z^{b}$, where the extended Lagrange (antisymmetric) matrix $\omega_{ab} \equiv \partial\Gamma_{b}/\partial z^{a} - \partial\Gamma_{a}/\partial z^{b}$ is constructed from the symplectic one-form $\Gamma$. Note that the Hamiltonian dynamics takes place on the surface ${\cal H} \equiv 0$, so that the energy coordinate is defined to be equal to the regular Hamiltonian: $w \equiv H({\bf x},{\bf p}, t)$. The inverse Lagrange matrix (assuming it is regular) yields the extended Poisson matrix $J^{ab} \equiv \{ z^{a},\; z^{b}\}$, in terms of which we express the extended Hamilton's equations $dz^{a}/\partial\sigma \equiv \{ z^{a},\; {\cal H}\} = J^{ab}\;\partial{\cal H}/\partial z^{b}$. Lastly, we note that an arbitrary exact exterior derivative $\exd S$ may be added to the phase-space Lagrangian $\Gamma - {\cal H}\,\exd\sigma$ without changing the equations of motion. 

When perturbations are introduced in the extended phase-space Lagrangian
\begin{equation}
\Gamma \;-\; {\cal H}\,\exd\sigma \;\equiv\; \left( \Gamma_{0} \;+\; \epsilon\,\Gamma_{1} \;+\frac{}{} \cdots \right) \;-\; \left( H_{0} \;+\; \epsilon\,H_{1} \;+\frac{}{} \cdots \;-\; w \right)\,\exd\sigma,
\label{eq:PSL}
\end{equation}
both the Poisson-bracket structure $\{\;,\;\}$ and the Hamiltonian ${\cal H}$ are perturbed. In the Hamiltonian formulation of Lie-transform phase-space Lagrangian perturbation theory, we search for a phase-space transformation $z \rightarrow \ov{z}$ (which induces a push-forward operation 
${\sf T}_{\epsilon}^{-1}$) such that the new symplectic one-form $\ov{\Gamma} \equiv {\sf T}_{\epsilon}^{-1}\Gamma + \exd{\cal S} \equiv \ov{\Gamma}_{0}$ contains no perturbation trace, while all perturbation effects are included in the new Hamiltonian $\ov{{\cal H}} \equiv {\sf T}_{\epsilon}^{-1}{\cal H} = \ov{{\cal H}}_{0} + \epsilon\,\ov{{\cal H}}_{1} + \epsilon^{2}\,\ov{{\cal H}}_{2} + \cdots$, where the second-order ponderomotive Hamiltonian 
$\ov{{\cal H}}_{2}$ contains nonlinear polarization and magnetization effects \cite{Brizard_Vlasovia} (generated by the reduced displacement 
${\sf T}_{\epsilon}^{-1}{\bf x}$).

\subsection{Unperturbed guiding-center Hamiltonian dynamics}

Modern gyrokinetic theory \cite{Brizard_Hahm} is based on a sequence of two near-identity phase-space transformations. The first transformation involves the time-independent transformation (in eight-dimensional extended phase space) from particle coordinates $z^{a} = ({\bf x}, {\bf p}; w,t)$ to guiding-center (gc) coordinates $z_{\rm gc}^{a} = ({\bf R}_{\rm gc}, p_{\|{\rm gc}}, \mu_{\rm gc}, \theta_{\rm gc}; W_{\rm gc},t)$. Here, the extended guiding-center Hamiltonian and the guiding-center symplectic one-form, which are respectively defined as
\begin{equation}
\left. \begin{array}{rcl}
{\cal H}_{0{\rm gc}} & \equiv & H_{0{\rm gc}} - W_{\rm gc} \;\equiv\; \mu_{\rm gc}\,B_{0} + p_{\|{\rm gc}}^{2}/2m - W_{\rm gc} \\
 &  & \\
\Gamma_{0{\rm gc}} & \equiv & [(e/c)\,{\bf A}_{0} + p_{\|{\rm gc}}\,\bhat_{0}]\bdot \exd{\bf R}_{\rm gc} + \mu_{\rm gc}\,(B_{0}/\Omega)\,\exd
\theta_{\rm gc} - W_{\rm gc}\,\exd t
\end{array} \right\}
\label{eq:gc_unperturbed}
\end{equation}
are independent of the guiding-center gyroangle $\theta_{\rm gc}$, and thus the guiding-center magnetic moment $\mu_{\rm gc}$ (or the gyroaction 
$\mu_{\rm gc}B_{0}/\Omega$ canonically conjugate to $\theta_{\rm gc}$) is an adiabatic invariant for the guiding-center motion of a charged particle in the time-independent magnetic field ${\bf B}_{0} \equiv \nabla\btimes{\bf A}_{0}$. In addition, the particle's energy coordinate $w$ and the guiding-center's energy coordinate $W_{\rm gc}$ are equal (because this transformation is time-independent). 

The guiding-center transformation introduces the push-forward operator ${\sf T}_{\rm gc}^{-1}$, which transforms functions $f$ on particle phase space to functions $f_{\rm gc} \equiv {\sf T}_{\rm gc}^{-1}f$ on guiding-center phase space, such that the scalar-invariance property $f(z) \equiv f_{\rm gc}(
z_{\rm gc})$ is satisfied. Here, we introduce two important applications of the guiding-center push-forward ${\sf T}_{\rm gc}^{-1}$. First, we introduce the guiding-center gyroradius 
\begin{equation}
\vb{\rho}_{\rm gc} \;\equiv\; {\sf T}_{\rm gc}^{-1}{\bf x} \;-\; {\bf R}_{\rm gc} \;=\; \vb{\rho}_{0{\rm gc}} \;+\; \epsilon_{\rm B}\;
\vb{\rho}_{1{\rm gc}} \;+\; \cdots,
\label{eq:rho_gc_def}
\end{equation}
which is expanded in powers of the guiding-center small parameter $\epsilon_{\rm B} \equiv \rho_{\rm th}/L_{\rm B} \ll 1$ (defined in terms of the thermal gyroradius $\rho_{\rm th} \equiv v_{\rm th}/|\Omega|$ and the characteristic nonuniformity length scale $L_{\rm B}$ of the background magnetic field). In Eq.~\eqref{eq:rho_gc_def}, the lowest-order gyroradius $\vb{\rho}_{0{\rm gc}}$ depends explicitly on the guiding-center gyroangle 
$\theta_{\rm gc}$, while the first-order correction $\vb{\rho}_{1{\rm gc}}$ has gyroangle-independent and gyroangle-dependent parts \cite{vgc_def}. The guiding-center transformation, therefore, introduces an electric-dipole $\vb{\pi}_{\rm gc} \equiv e\,\langle\vb{\rho}_{\rm gc}\rangle$ and an intrinsic (M) magnetic-dipole moment $\vb{\mu}_{\rm gc}^{M} \equiv (e/2c)\,\langle \vb{\rho}_{\rm gc}\btimes d_{\rm gc}\vb{\rho}_{\rm gc}/dt\rangle$ into the guiding-center Maxwell's equations. The lowest-order electric and magnetic dipole moments are \cite{Kaufman_gc}
\begin{equation} 
\left( \begin{array}{c}
\vb{\pi}_{1{\rm gc}} \\
\\
\vb{\mu}_{0{\rm gc}}^{\rm M}
\end{array} \right) \;\equiv\; \left( \begin{array}{c}
e\,\langle\vb{\rho}_{1{\rm gc}}\rangle \\
\\
(e\Omega/2c)\,\langle \vb{\rho}_{0{\rm gc}}\btimes \partial\vb{\rho}_{0{\rm gc}}/\partial\theta_{\rm gc}\rangle
\end{array} \right) \;=\; \left( \begin{array}{c}
e\,\bhat_{0}\btimes\langle{\bf v}_{1{\rm gc}}\rangle/\Omega \\
\\
-\,\mu_{\rm gc}\;\bhat_{0}
\end{array} \right),
\label{eq:pimu_gc}
\end{equation}
where $\langle{\bf v}_{1{\rm gc}}\rangle \equiv \dot{{\bf R}}_{{\rm gc}\bot}$ denotes the standard guiding-center magnetic-drift velocity. Note that the intrinsic magnetic-dipole moment \eqref{eq:pimu_gc} satisfies the variational derivation $\vb{\mu}_{0{\rm gc}}^{\rm M} \equiv -\;\partial 
H_{0{\rm gc}}/\partial{\bf B}_{0}$. In addition, the first-order guiding-center magnetic-dipole moment 
\begin{equation}
\vb{\mu}_{1{\rm gc}} \;\equiv\; \frac{e}{c} \left( \left\langle\wt{\vb{\rho}}_{1{\rm gc}}\btimes \Omega\;\pd{\vb{\rho}_{0{\rm gc}}}{\theta_{\rm gc}}\right\rangle \;+\; \langle\vb{\rho}_{1{\rm gc}}\rangle \btimes \dot{{\bf R}}_{{\rm gc}\|} \right)
\label{eq:mu1_gc}
\end{equation}
includes an intrinsic-magnetic-dipole contribution (involving the gyroangle-dependent part $\wt{\vb{\rho}}_{1{\rm gc}}$) and a moving-electric-dipole contribution (involving the gyroangle-independent part $\langle\vb{\rho}_{1{\rm gc}}\rangle$), with $\dot{{\bf R}}_{{\rm gc}\|} \equiv p_{\|{\rm gc}}\,\bhat_{0}/m$.

Next, we introduce the particle guiding-center velocity defined as the guiding-center push-forward of the particle velocity \cite{vgc_def}
\begin{equation}
{\bf v}_{\rm gc} \;\equiv\; {\sf T}_{\rm gc}^{-1}{\bf v} \;\equiv\; \{ {\bf R}_{\rm gc} + \vb{\rho}_{\rm gc},\; {\cal H}_{\rm gc}\}_{\rm gc} \;=\;
{\bf v}_{0{\rm gc}} + \epsilon_{\rm B}\,{\bf v}_{1{\rm gc}} + \cdots, 
\label{eq:vgc_def}
\end{equation} 
where ${\bf v}_{0{\rm gc}} = \dot{{\bf R}}_{{\rm gc}\|} + \Omega\,\partial\vb{\rho}_{0{\rm gc}}/\partial\theta_{\rm gc}$ and ${\bf v}_{1{\rm gc}} = 
\dot{{\bf R}}_{{\rm gc}\bot} + \wt{{\bf v}}_{1{\rm gc}}$. We note that the particle guiding-center velocity \eqref{eq:vgc_def} satisfies the relation
\begin{equation}
{\sf T}_{\rm gc}^{-1}\left(\frac{m}{2}\;|{\bf v}|^{2} \right) \;\equiv\; \frac{m}{2}\,|{\bf v}_{\rm gc}|^{2} \;=\; \mu_{\rm gc}\,B_{0} \;+\; 
\frac{p_{\|{\rm gc}}^{2}}{2m} \;+\; {\cal O}(\epsilon_{\rm B}^{2}), 
\label{eq:vgc_2}
\end{equation}
i.e., the guiding-center kinetic energy is identical to the particle kinetic energy (and thus the energy coordinates are identical: $W_{\rm gc} \equiv 
w$). 

\subsection{Perturbed guiding-center Hamiltonian dynamics}

When the perturbed electromagnetic potentials $A_{1}^{\nu} = (\phi_{1}, {\bf A}_{1})$ are introduced into the guiding-center phase-space Lagrangian formulation \eqref{eq:gc_unperturbed}, the perturbed guiding-center Hamiltonian $H_{1{\rm gc}}$ and the perturbed guiding-center symplectic one-form 
$\Gamma_{1{\rm gc}}$ are defined in Eq.~\eqref{eq:PSL} as \cite{Brizard_89}
\begin{equation}
\left. \begin{array}{rcl}
H_{1{\rm gc}} & \equiv & e\,\phi_{1{\rm gc}} \;\equiv\; {\sf T}_{\rm gc}^{-1}\left(e\frac{}{}\phi_{1}\right) \\
 &  & \\
\Gamma_{1{\rm gc}} & \equiv & (e/c)\,{\bf A}_{1{\rm gc}}\bdot\exd({\bf R}_{\rm gc} + \vb{\rho}_{\rm gc}) \;\equiv\; {\sf T}_{\rm gc}^{-1}\left[ 
(e/c)\frac{}{}{\bf A}_{1}\bdot\exd{\bf x} \right]
\end{array} \right\}.
\label{eq:gc_perturbed}
\end{equation}
Here, the electromagnetic potentials $A_{1}^{\nu}$ are transformed by the guiding-center push-forward operator ${\sf T}_{\rm gc}^{-1}$ into the guiding-center potentials $A_{1{\rm gc}}^{\nu} \equiv (\phi_{1{\rm gc}}, {\bf A}_{1{\rm gc}})$, where
\begin{equation}
A_{1{\rm gc}}^{\nu} \;\equiv\; {\sf T}_{\rm gc}^{-1}\;A_{1}^{\nu} \;\equiv\; A_{1}^{\nu}({\bf R}_{\rm gc} + \vb{\rho}_{\rm gc}, t).
\label{eq:gcpush_phiA}
\end{equation}
The appearance of gyroangle dependence in the perturbed guiding-center formulation \eqref{eq:gc_perturbed} implies that the guiding-center magnetic moment $\mu_{\rm gc}$ is no longer conserved, i.e., $\dot{\mu}_{\rm gc} = -\,\epsilon\,(e\Omega/B_{0})\,\partial\phi_{1{\rm gc}}/\partial\theta_{\rm gc}
+ \cdots$, where the small ordering parameter $\epsilon$ denotes the strength of the field fluctuations.

\subsection{Gyrocenter phase-space transformation}

In order to restore the (adiabatic) invariance of the magnetic moment $\mu$, a second phase-space transformation is introduced after the guiding-center transformation. The gyrocenter transformation involves the time-dependent transformation from guiding-center coordinates $z_{\rm gc}^{a}$ to the gyrocenter (gy) coordinates $Z^{a} = ({\bf R}, p_{\|}, \mu, \theta; W,t)$:
\begin{equation}
Z^{a} \;\equiv\; z_{\rm gc}^{a} \;+\; \epsilon\;G_{1}^{a} \;+\; \epsilon^{2} \left( G_{2}^{a} \;+\; \frac{1}{2}\;G_{1}^{b}\;
\pd{G_{1}^{a}}{z_{\rm gc}^{b}} \right) \;+\; \cdots,
\label{eq:ZZ_0}
\end{equation}
which is generated by the phase-space vector-components $(G_{1}^{a}, G_{2}^{a}, \cdots)$. Through this transformation, the fast gyromotion is once again (asymptotically) decoupled from the slow gyrocenter Hamiltonian dynamics (based on the time-scale separation between the fast gyromotion and the characteristic fluctuation time scale) and the gyrocenter magnetic moment $\mu$ is now an adiabatic invariant for the gyrocenter motion of a charged particle in the perturbed electromagnetic fields $\epsilon\,{\bf E}_{1}$ and ${\bf B}_{0} + \epsilon\,{\bf B}_{1}$. 

The gyrocenter transformation \eqref{eq:ZZ_0} induces a transformation on the perturbed guiding-center phase-space Lagrangian formulation \eqref{eq:gc_perturbed}, generated by the gyrocenter push-forward operator ${\sf T}_{\rm gy}^{-1}$, such that the new gyrocenter Hamiltonian dynamics is expressed in terms of the gyrocenter Hamiltonian and the gyrocenter symplectic one-form:
\begin{equation}
\left. \begin{array}{rcl}
{\cal H}_{\rm gy} & \equiv & {\sf T}_{\rm gy}^{-1}{\cal H}_{\rm gc} \;\equiv\; H_{\rm gy} \;-\; W \\
 &  & \\
\Gamma_{\rm gy} & \equiv & {\sf T}_{\rm gy}^{-1}\Gamma_{\rm gc} \;+\; \exd {\cal S}_{\rm gy}
\end{array} \right\},
\label{eq:gy_dynamics}
\end{equation}
where the phase-space gauge function ${\cal S}_{\rm gy} \equiv \epsilon\,S_{1} + \epsilon^{2}\,S_{2} + \cdots$ (chosen to be explicitly gyroangle-dependent at all orders in $\epsilon$) generates the canonical part of the phase-space transformation \eqref{eq:ZZ_0} [see 
Eqs.~\eqref{eq:G1_def}-\eqref{eq:G2_def} below]. By construction, the gyroangle-component of the gyrocenter symplectic form $\Gamma_{\rm gy}$ is $(mc/e)\,\mu\,\exd\theta$ and the gyrocenter Hamiltonian $H_{\rm gy}$ is independent of the gyrocenter gyroangle $\theta$.

In the present work, we use the Hamiltonian formulation of gyrokinetic theory \cite{Brizard_Hahm}, where the magnetic perturbation term ${\bf A}_{1}$ is removed from the perturbed guiding-center phase-space Lagrangian formulation \eqref{eq:gc_perturbed}, so that the gyrocenter symplectic one-form is simply defined as $\Gamma_{\rm gy} \equiv \Gamma_{0{\rm gc}}$ (i.e., the gyrocenter Poisson bracket is identical to the unperturbed guiding-center Poisson bracket, denoted as $\{\;,\;\}_{\rm gc}$). The components of the phase-space generating vector fields $({\sf G}_{1}, {\sf G}_{2}, \cdots)$ that achieve this transformation of the symplectic one-form are \cite{Brizard_89}
\begin{eqnarray}
G_{1}^{a} & \equiv & \{ S_{1},\; z_{\rm gc}^{a} \}_{\rm gc} \;+\; \frac{e}{c}\,{\bf A}_{1{\rm gc}}\bdot\{ {\bf R}_{\rm gc} + \vb{\rho}_{\rm gc},\; 
z_{\rm gc}^{a}\}_{\rm gc},
\label{eq:G1_def} \\
G_{2}^{a} & \equiv & \{ S_{2},\; z_{\rm gc}^{a} \}_{\rm gc} \;-\; \frac{e}{2c}\,\left\{ {\bf R}_{\rm gc} + \vb{\rho}_{\rm gc},\frac{}{} S_{1} 
\right\}_{\rm gc}\btimes{\bf B}_{1{\rm gc}}\bdot\{ {\bf R}_{\rm gc} + \vb{\rho}_{\rm gc},\; z_{\rm gc}^{a}\}_{\rm gc},
\label{eq:G2_def}
\end{eqnarray}
where the time-invariance condition $G_{n}^{t} \equiv 0$ (at all orders $n \geq 1$) implies that $\partial S_{n}/\partial w \equiv 0$. We note that the gyrocenter Jacobian associated with the phase-space transformation \eqref{eq:ZZ_0} is defined as
\begin{equation}
{\cal J}_{\rm gy} \;\equiv\; {\cal J}_{\rm gc} \;-\; \epsilon\;\partial_{a}\left({\cal J}_{\rm gc}\frac{}{}G_{1}^{a}\right) \;+\; \cdots \;=\; 
{\cal J}_{0{\rm gc}} \;+\; \epsilon \left[\; {\cal J}_{1{\rm gc}} \;-\; \partial_{a}\left({\cal J}_{0{\rm gc}}\frac{}{}G_{1}^{a}\right) \;\right] 
\;+\; \cdots \;\equiv\; {\cal J}_{0{\rm gc}},
\label{eq:gy_Jac}
\end{equation}
where the unperturbed guiding-center Jacobian is ${\cal J}_{0{\rm gc}} \equiv \bhat_{0}\bdot\nabla\btimes({\bf A}_{0} + p_{\|{\rm gc}}\,c\,\bhat_{0}/e)$ and the first-order perturbed guiding-center Jacobian is ${\cal J}_{1{\rm gc}} \equiv \bhat_{0}\bdot\nabla\btimes{\bf A}_{1{\rm gc}}$. Within the Hamiltonian formulation of gyrokinetic theory, the identity ${\cal J}_{\rm gy} \equiv {\cal J}_{0{\rm gc}}$ in Eq.~\eqref{eq:gy_Jac} is thus guaranteed by the identity ${\cal J}_{1{\rm gc}} \equiv \partial_{a}({\cal J}_{0{\rm gc}}\,G_{1}^{a})$, where we used the canonical identity 
$\partial_{a}({\cal J}_{0{\rm gc}}\,\{S_{1},\; z_{\rm gc}^{a}\}_{\rm gc}) \equiv 0$. 

The gauge functions $(S_{1}, S_{2}, \cdots)$ in Eqs.~\eqref{eq:G1_def}-\eqref{eq:G2_def} are chosen to eliminate the gyroangle-dependence in the perturbed guiding-center Hamiltonian (at arbitrary order in the perturbation analysis). The gyrocenter Hamiltonian in Eq.~\eqref{eq:gy_dynamics} is thus defined (up to $\epsilon^{2}$) as 
\begin{equation}
H_{\rm gy} \;\equiv\; H_{0{\rm gy}} \;+\; \epsilon\;e\,\left\langle\psi_{1{\rm gc}}\right\rangle \;+\; \frac{\epsilon^{2}}{2} \left(
\frac{e^{2}}{mc^{2}}\; \left\langle|{\bf A}_{1{\rm gc}}|^{2}\right\rangle \;-\; e\;\left\langle \left\{ S_{1},\frac{}{}
\wt{\psi}_{1{\rm gc}} \right\}_{\rm gc} \right\rangle \right),
\label{eq:Hgy_def}
\end{equation}
where the zeroth-order term is simply the unperturbed guiding-center Hamiltonian $H_{0{\rm gy}} \equiv \mu\,B_{0} + p_{\|}^{2}/2m$, while the first-order (covariant) guiding-center effective potential
\begin{equation} 
\psi_{1{\rm gc}} \;\equiv\; {\sf T}_{\rm gc}^{-1}\left(\phi_{1} \;-\; {\bf A}_{1}\bdot\frac{{\bf v}}{c} \right) \;\equiv\; \phi_{1{\rm gc}} \;-\; 
{\bf A}_{1{\rm gc}}\bdot\frac{{\bf v}_{\rm gc}}{c},
\label{eq:psi_def}
\end{equation}
is expressed in terms of the guiding-center potentials \eqref{eq:gcpush_phiA} and the guiding-center particle velocity \eqref{eq:vgc_def}. Lastly, the first-order function $S_{1}$ in Eq.~\eqref{eq:Hgy_def} is formally defined as the solution of the first-order equation 
\begin{equation}
\frac{d_{\rm gc}S_{1}}{dt} \;\equiv\; \{ S_{1},\; {\cal H}_{0{\rm gc}}\}_{\rm gc} \;=\; e\,\wt{\psi}_{1{\rm gc}}, 
\label{eq:S1_ODE}
\end{equation}
which is approximated (to lowest order in the gyrokinetic space-time-scale ordering \cite{Brizard_Hahm}) as
\begin{equation}
S_{1} \;\equiv\; \left( \frac{d_{\rm gc}}{dt}\right)^{-1}\,e\,\wt{\psi}_{1{\rm gc}} \;\simeq\; \frac{e}{\Omega}\;\int\,\wt{\psi}_{1{\rm gc}}\;d\theta.
\label{eq:S1_def}
\end{equation}
We note that, while the gyroangle-independent potential $\langle\psi_{1{\rm gc}}\rangle$ contributes to the linear (first-order) perturbed gyrocenter Hamiltonian dynamics, the gyroangle-dependent potential $\wt{\psi}_{1{\rm gc}}$ contributes to the (second-order) gyrocenter ponderomotive Hamiltonian in Eq.~\eqref{eq:Hgy_def}. In addition, we note that the appearance of the magnetic term ${\bf A}_{1}$ in Eq.~\eqref{eq:Hgy_def} is a reflection of the Hamiltonian formulation adopted in the present work. We now show in the next Section that the ponderomotive effects (associated with 
$\wt{\psi}_{1{\rm gc}}$) are also related to polarization and magnetization effects that appear in the gyrokinetic Maxwell equations.

\section{\label{sec:gyro_polmag}Gyrocenter Polarization and Magnetization}

The gyrocenter push-forward operator ${\sf T}_{\rm gy}^{-1}$ and the pull-back operator ${\sf T}_{\rm gy}$ introduced by the gyrocenter phase-space transformation \eqref{eq:ZZ_0} play a fundamental role in modern gyrokinetic theory \cite{Brizard_Hahm}. While the gyrocenter pull-back operator 
${\sf T}_{\rm gy}$ is involved in the lowest-order integration of the gyrokinetic Vlasov equation \cite{Briz_Mish}, the push-forward operator 
${\sf T}_{\rm gy}^{-1}$ is intimately connected with the reduced polarization and magnetization effects \cite{Brizard_JCPS,Brizard_Vlasovia} in the gyrokinetic Maxwell equations. 

\subsection{Gyrocenter Gyroradius}

Reduced polarization and magnetization effects in nonlinear gyrokinetic theory are associated with the first-order gyrocenter gyroradius 
$\vb{\rho}_{1{\rm gy}}$, defined as the gyrocenter push-forward of the particle guiding-center position
\begin{eqnarray}
\epsilon\;\vb{\rho}_{1{\rm gy}} \;+\; \cdots & \equiv & {\sf T}_{\rm gy}^{-1}\left({\sf T}_{\rm gc}^{-1}{\bf x}\right) \;-\; {\sf T}_{\rm gc}^{-1}{\bf x}
\;=\; \epsilon\;\left\{ {\bf R} \;+\; \vb{\rho}_{0},\frac{}{} S_{1} \right\}_{\rm gc} \;+\; \cdots \nonumber \\
 & = & \epsilon\;\left[ \frac{\Omega}{B_{0}} \left( \pd{\vb{\rho}_{0}}{\theta}\;\pd{S_{1}}{\mu} \;-\; \pd{\vb{\rho}_{0}}{\mu}\;
\pd{S_{1}}{\theta} \right) \;+\; \bhat_{0}\;\pd{S_{1}}{p_{\|}} \;+\; \frac{c\bhat_{0}}{eB_{0}}\btimes\nabla S_{1} \right] \;+\; \cdots,
\label{eq:rhogy_def}
\end{eqnarray}
which is expressed exclusively in terms of $S_{1}$ (and therefore $\wt{\psi}_{1{\rm gc}}$) and we henceforth use the lowest-order guiding-center gyroradius $\vb{\rho}_{0} \equiv \vb{\rho}_{0{\rm gc}}$ and guiding-center particle velocity ${\bf v}_{0} \equiv {\bf v}_{0{\rm gc}}$. 

The first-order gyrocenter gyroradius \eqref{eq:rhogy_def} can be decomposed into a gyroangle-independent part 
\begin{equation}
\langle\vb{\rho}_{1{\rm gy}}\rangle \;\equiv\; -\;\frac{\Omega}{B_{0}}\;\pd{}{\mu}\left\langle \vb{\rho}_{0}\;\pd{S_{1}}{\theta} \right\rangle \;=\;
-\;\frac{e}{B_{0}}\;\pd{}{\mu}\left\langle \vb{\rho}_{0}\frac{}{}\wt{\psi}_{1{\rm gc}}\right\rangle,
\label{eq:rhogy_ave}
\end{equation}
and a gyroangle-dependent part 
\begin{equation}
\wt{\vb{\rho}}_{1{\rm gy}} \;\equiv\; \vb{\rho}_{1{\rm gy}} \;-\; \langle\vb{\rho}_{1{\rm gy}}\rangle \;=\; \bhat_{0}\;\pd{S_{1}}{p_{\|}} \;+\;
\frac{c\bhat_{0}}{eB_{0}}\btimes\nabla S_{1} \;+\; \frac{\Omega}{B_{0}} \left[\; \pd{}{\theta} \left( \vb{\rho}_{0}\;\pd{S_{1}}{\mu} \right) \;-\;
\pd{}{\mu} \left( \vb{\rho}_{0}\;\pd{S_{1}}{\theta} \;-\; \left\langle \vb{\rho}_{0}\;\pd{S_{1}}{\theta} \right\rangle \right) \;\right].
\label{eq:rhogy_dep}
\end{equation} 
We note that the gyroangle-independent part \eqref{eq:rhogy_ave} has only perpendicular components while the gyroangle-dependent part 
\eqref{eq:rhogy_dep} has both parallel and perpendicular components. 

\subsection{Gyrocenter Dipole Moments}

We now make a few comments concerning the role played by the first-order gyrocenter gyroradius \eqref{eq:rhogy_def} in modern gyrokinetic theory. First, according to the general reduced Vlasov-Maxwell theory \cite{Brizard_Vlasovia}, the gyrocenter transformation introduces additional electric and magnetic dipole moments (for each particle species), which are defined as 
\begin{eqnarray}
\vb{\pi}_{1{\rm gy}} & \equiv & e\;\langle\vb{\rho}_{1{\rm gy}}\rangle \;=\; -\;\frac{e^{2}}{B_{0}}\;\pd{}{\mu}\left\langle \vb{\rho}_{0}
\frac{}{}\wt{\psi}_{1{\rm gc}}\right\rangle, \label{eq:pigy_def} \\
\vb{\mu}_{1{\rm gy}} & \equiv & e \left\langle \vb{\rho}_{1{\rm gy}}\btimes\frac{{\bf v}_{0}}{c} \right\rangle \;=\; \frac{e}{c} \left\langle 
\wt{\vb{\rho}}_{1{\rm gy}}\btimes\Omega\,\pd{\vb{\rho}_{0}}{\theta} \right\rangle \;+\; e\,\langle\vb{\rho}_{1{\rm gy}}\rangle\btimes \frac{p_{\|}\,
\bhat_{0}}{mc} \nonumber \\
 & = & \frac{e\,\bhat_{0}}{B_{0}} \left\langle\vb{\rho}_{0}\bdot\nabla\frac{}{}\wt{\psi}_{1{\rm gc}} \right\rangle \;+\; \frac{e^{2}}{c} 
\left\langle\pd{\vb{\rho}_{0}}{\theta}\;\pd{\wt{\psi}_{1{\rm gc}}}{p_{\|}} \right\rangle \;-\;\frac{e^{2}\,p_{\|}}{mc\,B_{0}}\;\pd{}{\mu}
\left\langle \pd{\vb{\rho}_{0}}{\theta}\;\wt{\psi}_{1{\rm gc}} \right\rangle.
\label{eq:mugy_def}
\end{eqnarray}
Here, we note that the magnetic-dipole moment \eqref{eq:mugy_def} is naturally decomposed \cite{Jackson,moving_mag} into an intrinsic part (involving 
$\wt{\vb{\rho}}_{1{\rm gy}}$) and a moving-electric-dipole part (involving $\langle\vb{\rho}_{1{\rm gy}}\rangle$). The gyrocenter definitions 
\eqref{eq:pigy_def}-\eqref{eq:mugy_def} are therefore natural extensions of the guiding-center electric and magnetic dipole moments \eqref{eq:pimu_gc}. 

In order to get a clearer picture of these gyrocenter dipole moments, we apply the zero-Larmor-radius (ZLR) limit $(\vb{\rho}_{0} \rightarrow 0)$ on the gyrocenter moments \eqref{eq:pigy_def}-\eqref{eq:mugy_def}. In this limit, the first-order gyrocenter electric-dipole and intrinsic magnetic-dipole moments are $\vb{\pi}_{1{\rm gy}} \simeq e\;\vb{\rho}_{1\bot}$ and $\vb{\mu}_{1{\rm gy}}^{\rm M} \simeq -\,\mu\,{\bf B}_{1}/B_{0}$, where the first-order gyrocenter gyroradius (in the ZLR limit) is
\begin{equation} 
\vb{\rho}_{1\bot} \;\equiv\; \frac{e}{m\Omega^{2}}\;\left( {\bf E}_{1\bot} \;+\; \frac{p_{\|}\bhat_{0}}{mc}\btimes{\bf B}_{1\bot} \right) \;+\; 
\bhat_{0}\btimes\;\frac{{\bf A}_{1\bot}}{B_{0}}.
\label{eq:rho1_bot}
\end{equation}
Hence, in the ZLR gyrofluid approximation \cite{Brizard_92}, the gyrofluid polarization charge density is expressed (at lowest order) as
\begin{equation}
-\;\nabla\bdot \left(e\,n\frac{}{}\vb{\rho}_{1\bot}\right) \;=\; -\;\nabla\left[ \frac{mn\,c^{2}}{B_{0}^{2}} \left( {\bf E}_{1\bot} \;+\; \frac{u_{\|}\bhat_{0}}{c}\btimes{\bf B}_{1\bot} \right) \right] \;+\; e\,n\;\frac{B_{1\|}}{B_{0}},
\label{eq:gyro_pol}
\end{equation}
where $n$ and $u_{\|}$ denote the gyrofluid density and parallel gyrofluid velocity. In Eq.~\eqref{eq:gyro_pol}, we easily recognize the perpendicular  perturbed electric and magnetic-flutter contributions as well as the compressional contribution from the parallel perturbed magnetic field. On the other hand, the ZLR gyrocenter polarization velocity
\[ \pd{\vb{\rho}_{1\bot}}{t} \;=\; \frac{c}{B_{0}\Omega} \left( \pd{{\bf E}_{1\bot}}{t} \;+\; \frac{p_{\|}\bhat_{0}}{mc}\btimes\pd{{\bf B}_{1\bot}}{t}
\right) \;-\; \left( \frac{1}{c}\;\pd{{\bf A}_{1\bot}}{t}\right)\btimes\frac{c\bhat_{0}}{B_{0}}, \]
includes the standard polarization velocity (with its magnetic-flutter counter-part) and the inductive part of the $E\times B$ velocity, i.e.,
$({\bf E}_{1} + \nabla\phi_{1})\btimes c\bhat_{0}/B_{0} \equiv -\,\partial{\bf A}_{1}/\partial t \btimes\bhat_{0}/B_{0}$.

\subsection{Gyrocenter Push-forward of the Guiding-center Particle Velocity}

Next, we consider the gyrocenter push-forward of the guiding-center velocity
\begin{eqnarray}
{\sf T}_{\rm gy}^{-1}{\bf v}_{0} & = & {\bf v}_{0} \;-\; \epsilon \left( \left\{ S_{1},\frac{}{} {\bf v}_{0} \right\}_{\rm gc} \;+\;
\frac{e\,{\bf A}_{1{\rm gc}}}{mc} \right) \;+\; \cdots \nonumber \\
 & \equiv & \left\{ {\bf R} + \vb{\rho}_{0},\frac{}{} \left( H_{0{\rm gc}} \;+\; \epsilon\;e\,\langle\psi_{1{\rm gc}}\rangle \;+\; \cdots \right)
\right\}_{\rm gc} \;+\; \epsilon\;\frac{d_{\rm gc}}{dt}\,\vb{\rho}_{1{\rm gy}} \;+\; \cdots,
\label{eq:push_velocity}
\end{eqnarray}
where we used the relation
\[ -\;\left\{ S_{1},\frac{}{} {\bf v}_{0} \right\}_{\rm gc} \;\equiv\; \frac{d_{\rm gc}}{dt}\,\vb{\rho}_{1{\rm gy}} \;+\; \left\{ {\bf R} \;+\frac{}{} 
\vb{\rho}_{0},\;e\,\langle\psi_{1{\rm gc}}\rangle \right\}_{\rm gc} \;+\; \frac{e\,{\bf A}_{1{\rm gc}}}{mc}, \]
with the total first-order polarization-displacement velocity \cite{Brizard_Vlasovia} defined as 
\begin{equation}
\frac{d_{\rm gc}\vb{\rho}_{1{\rm gy}}}{dt} \;\equiv \; \left\{ \vb{\rho}_{1{\rm gy}},\frac{}{} {\cal H}_{0{\rm gc}} \right\}_{\rm gc} \;=\; 
\pd{\vb{\rho}_{1{\rm gy}}}{t} \;+\; \left\{ \vb{\rho}_{1{\rm gy}},\frac{}{} H_{0{\rm gc}} \right\}_{\rm gc}. 
\label{eq:drho1_dt}
\end{equation}
The gyroangle average of the gyrocenter push-forward relation \eqref{eq:push_velocity} is expressed as
\begin{equation}
\left\langle{\sf T}_{\rm gy}^{-1}{\bf v}_{0}\right\rangle \;=\; \left\{ {\bf R},\frac{}{} \left( H_{0{\rm gc}} + \epsilon\;e\,\langle
\psi_{1{\rm gc}}\rangle \;+\frac{}{} \cdots \right) \right\}_{\rm gc} \;+\; \left( \epsilon \frac{d_{\rm gc}}{dt}\,\langle\vb{\rho}_{1{\rm gy}}\rangle 
\;+\; \cdots \right),
\label{eq:vgy_average}
\end{equation}
which displays the characteristic separation \cite{Sosenko,Brizard_Vlasovia} of the gyrocenter velocity $\dot{{\bf R}} \equiv \dot{{\bf R}}_{\rm gc} +
\epsilon\,\dot{{\bf R}}_{1{\rm gy}} + \cdots$, where 
\begin{equation}
\dot{{\bf R}}_{1{\rm gy}} \;\equiv\; \left\{ {\bf R},\frac{}{} e\,\langle\psi_{1{\rm gc}}\rangle \right\}_{\rm gc} \;=\; -\;\frac{e\,\langle 
A_{1\|{\rm gc}}\rangle}{m\;c}\;\bhat_{0} \;+\; \frac{c\bhat_{0}}{B_{0}}\btimes\nabla\langle\psi_{1{\rm gc}}\rangle, 
\label{eq:R1dot}
\end{equation}
and the total first-order gyrocenter polarization velocity $d_{\rm gc}\langle\vb{\rho}_{\rm gy}\rangle/dt$. Note that, while the gyroangle-averaged effective potential $\langle\psi_{1{\rm gc}}\rangle$ contributes to the slow gyrocenter dynamics, the gyroangle-dependent potential 
$\wt{\psi}_{1{\rm gc}}$ contributes to the gyrocenter polarization velocity (through $\langle\vb{\rho}_{1{\rm gy}}\rangle$). 

\subsection{Polarization and Magnetization Effects in the Gyrocenter Ponderomotive Hamiltonian}

Lastly, we point out that polarization and magnetization effects associated with the first-order gyrocenter gyroradius \eqref{eq:rhogy_def} can also be made to appear explicitly in the gyrocenter Hamiltonian \eqref{eq:Hgy_def} by rewriting the second-order gyrocenter Hamiltonian as
\begin{equation}
H_{2{\rm gy}} \;=\; -\;\frac{e}{2c} \left\langle {\bf A}_{1{\rm gc}}\bdot\frac{}{}\left\{ {\bf R} + \vb{\rho}_{0},\frac{}{} e\,\left\langle
\psi_{1{\rm gc}} \right\rangle \right\}_{\rm gc} \right\rangle \;-\; \frac{e}{2}\; \left\langle \vb{\rho}_{1{\rm gy}}\bdot\left( {\bf E}_{1{\rm gc}}
\;+\; \frac{{\bf v}_{0}}{c}\btimes{\bf B}_{1{\rm gc}} \right) \right\rangle,
\label{eq:H2gy_rhogy}
\end{equation}
where we have defined the guiding-center perturbed electromagnetic fields
\begin{equation}
\left. \begin{array}{rcl}
{\bf E}_{1{\rm gc}} & \equiv & -\,\nabla\phi_{1{\rm gc}} \;-\; c^{-1}\partial{\bf A}_{1{\rm gc}}/\partial t \\
 &  & \\
{\bf B}_{1{\rm gc}} & \equiv & \nabla\btimes{\bf A}_{1{\rm gc}}
\end{array} \right\}. 
\label{eq:EB_1gc}
\end{equation}
We note that, within the Hamiltonian formulation of gyrokinetic theory, the ponderomotive term in Eq.~\eqref{eq:H2gy_rhogy} appears in the form of a 
{\it nonlinear} finite-Larmor-radius correction \cite{NFLR} as follows. 

First, we introduce the gyrocenter push-forward ${\sf T}_{\rm gy}^{-1}\psi_{1{\rm gc}}$ of the effective first-order potential \eqref{eq:psi_def}, defined as
\begin{eqnarray}
\psi_{1{\rm gy}} \;\equiv\; {\sf T}_{\rm gy}^{-1}\psi_{1{\rm gc}} & = & \psi_{1{\rm gc}} \;-\; \frac{\epsilon}{c}\;{\bf A}_{1{\rm gc}}\bdot
\left\{ {\bf R} + \vb{\rho}_{0},\frac{}{} e\,\left\langle\psi_{1{\rm gc}} \right\rangle \right\}_{\rm gc} \nonumber \\
 &  &-\; \epsilon\;\vb{\rho}_{1{\rm gy}}\bdot\left( {\bf E}_{1{\rm gc}} \;+\; \frac{{\bf v}_{0}}{c}\btimes{\bf B}_{1{\rm gc}} \right) \;-\;
\frac{d_{\rm gc}}{dt}\left( \frac{\epsilon}{c}\;{\bf A}_{1{\rm gc}}\bdot\vb{\rho}_{1{\rm gy}} \right) \;+\; \cdots,
\label{eq:psi1_gy}
\end{eqnarray}
where we used the definitions \eqref{eq:EB_1gc} for the guiding-center electromagnetic fields. Next, by replacing the guiding-center perturbation potential $\psi_{1{\rm gc}}$ with the gyrocenter perturbation potential \eqref{eq:psi1_gy} in the gyrocenter Hamiltonian and ignoring the exact time derivative in Eq.~\eqref{eq:psi1_gy}, since the Hamiltonian is defined up to an exact time derivative within Hamilton's variational principle, the gyrocenter Hamiltonian is now expressed as
\begin{equation}
H_{\rm gy} \;=\; H_{0{\rm gc}} \;+\; \epsilon\,e\;\langle\psi_{1{\rm gy}}\rangle \;+\; \frac{\epsilon^{2}}{2}\;e\;\left\langle 
\vb{\rho}_{1{\rm gy}}\bdot\left( {\bf E}_{1{\rm gc}}\;+\; \frac{{\bf v}_{0}}{c}\btimes{\bf B}_{1{\rm gc}} \right) \right\rangle,
\label{eq:Hgy_polmag}
\end{equation}
where the second term on the right side involves the gyroangle average of the gyrocenter potential \eqref{eq:psi1_gy}.

\section{\label{sec:full}Full Hamiltonian Gyrokinetic Formulation}

In this Section, we review the exact energy conservation law for the full-$f$ version of the nonlinear gyrokinetic Vlasov-Maxwell equations within the context of the gyrocenter polarization and magnetization effects introduced by the gyrocenter phase-space transformation \eqref{eq:ZZ_0}. The explicit proof of energy conservation is presented here with a focus on the role played by gyrocenter polarization and magnetization effects as they appear in both the gyrocenter Hamiltonian and the gyrokinetic Maxwell equations.

\subsection{Variational formulation}

The variational derivation of the full-$f$ version of the gyrokinetic Vlasov-Maxwell equations is based the gyrokinetic action functional 
\cite{Brizard_VP1,Brizard_VP2}
\begin{equation}
{\cal A}_{\rm gy} \;=\; -\;\int\;{\cal F}_{\rm gy}\;{\cal H}_{\rm gy}\;d^{8}Z \;+\; \int\frac{d^{4}x}{8\pi} \left( \epsilon^{2}\;|{\bf E}_{1}|^{2} 
\;-\frac{}{} |{\bf B}_{0} \;+\; \epsilon\;{\bf B}_{1}|^{2} \right),
\label{eq:gy_action}
\end{equation}
where the extended gyrocenter Hamiltonian ${\cal H}_{\rm gy} \equiv H_{\rm gy} - W$ is defined in Eq.~\eqref{eq:Hgy_def}, while the extended gyrocenter Vlasov distribution ${\cal F}_{\rm gy} \equiv F\;\delta(W - H_{\rm gy})$ guarantees that the energy constraint ${\cal H}_{\rm gy} \equiv 0$ is satisfied. 

In the full-$f$ version of the Hamiltonian formulation of gyrokinetic theory \cite{Brizard_Hahm}, the time evolution of $F$ is determined by the nonlinear gyrokinetic Vlasov equation
\begin{equation}
\pd{F}{t} \;+\; \left\{ F,\frac{}{} H_{\rm gy} \right\}_{\rm gc} \;=\; 0,
\label{eq:gy_Vlasov}
\end{equation}
where the gyrocenter Hamiltonian \eqref{eq:Hgy_def} is now written as
\begin{equation}
H_{\rm gy} \;\equiv\; \frac{p_{\|}^{2}}{2m} \;+\; \mu\,B_{0} \;+\; \epsilon\,e\;\langle \psi_{1{\rm gc}}\rangle \;-\; \frac{\epsilon^{2}}{2}\,e\;
\left\langle \pounds_{\rm gy}\,\psi_{1{\rm gc}}\right\rangle
\label{eq:gy_Hamiltonian}
\end{equation}
where the first-order gyrocenter Lie-derivative operator, generated by the first-order field \eqref{eq:G1_def}, is defined in terms of an arbitrary function $g$ as
\begin{equation}
\pounds_{\rm gy}\,g \;\equiv\; \left\{ S_{1},\frac{}{} g \right\}_{\rm gc} \;+\; \frac{e}{c}\,{\bf A}_{1{\rm gc}}\bdot\left\{ {\bf R} + 
\vb{\rho}_{0},\frac{}{} g \right\}_{\rm gc}.
\label{eq:L1_def}
\end{equation}
Next, we introduce the variational expression
\begin{equation}
\delta H_{\rm gy} \;=\; \epsilon\,e\;\langle \delta\psi_{1{\rm gc}}\rangle \;-\; \epsilon^{2}\,e\;\left\langle \pounds_{\rm gy}\,\delta\psi_{1{\rm gc}}\right\rangle \;\equiv\; \epsilon\,e\;\left\langle {\sf T}_{\epsilon}^{-1}\delta\psi_{1{\rm gc}}\right\rangle,
\label{eq:delta_Hgy}
\end{equation}
where the (truncated) gyrocenter push-forward operator ${\sf T}_{\epsilon}^{-1}$ is defined as
\begin{equation}
{\sf T}_{\epsilon}^{-1}g \;\equiv\; g \;-\; \epsilon\;\pounds_{\rm gy}\,g.
\label{eq:gy_push}
\end{equation} 
and we used the identity associated with the definition (\ref{eq:S1_def}) for $S_{1}$:
\begin{eqnarray} 
\left\langle\left\{ S_{1},\; e\,\delta\psi_{1{\rm gc}} \right\}_{\rm gc}\right\rangle & = & \left\langle\left\{ S_{1},\frac{}{} \{ \delta S_{1},\;
{\cal H}_{0{\rm gc}}\}_{\rm gc} \right\}_{\rm gc}\right\rangle \;\equiv\; \left\langle\left\{ \delta S_{1},\frac{}{} \{ S_{1},\;{\cal H}_{0{\rm gc}}
\}_{\rm gc} \right\}_{\rm gc}\right\rangle \;+\; \frac{d_{\rm gc}}{dt}\left\langle \left\{ S_{1},\; \delta S_{1}\right\}_{\rm gc}\right\rangle 
\nonumber \\
 & \equiv & \left\langle\left\{ \delta S_{1},\; e\,\psi_{1{\rm gc}} \right\}_{\rm gc}\right\rangle,
\label{eq:id_S1}
\end{eqnarray}
which is obtained by using the Jacobi identity for the guiding-center Poisson bracket and omitting the exact time derivative from the variation 
$\delta H_{\rm gy}$ (which is allowed in Hamiltonian dynamics). Equation (\ref{eq:delta_Hgy}) can also be used to write the functional variation of the gyrocenter Hamiltonian (\ref{eq:gy_Hamiltonian})
\begin{equation}
\delta H_{\rm gy} \;\equiv\; \int e\; \left[\; \left\langle {\sf T}_{\epsilon}^{-1}\delta_{\rm gc}^{3}\right\rangle\;\delta\phi_{1}({\bf r}) \;-\;
\left\langle {\sf T}_{\epsilon}^{-1}\left(\delta_{\rm gc}^{3}\;\frac{{\bf v}_{0}}{c}\right)\right\rangle\bdot\delta{\bf A}_{1}({\bf r}) \;\right]
\;d^{3}r,
\label{eq:delta_Hgy_phiA}
\end{equation}
where the guiding-center delta function 
\begin{equation}
\delta_{\rm gc}^{3} \;\equiv\; {\sf T}_{\rm gc}^{-1}\delta^{3}({\bf x} - {\bf r}) \;=\; \delta^{3}({\bf R} + \vb{\rho}_{0} - {\bf r}) 
\label{eq:delta_gc}
\end{equation}
is evaluated at an arbitrary field point ${\bf r}$ and the identity
\[ \delta\psi_{\rm gc} \;\equiv\; \int d^{3}r \left( \delta\phi_{1}({\bf r},t) \;-\; \frac{{\bf v}}{c}\bdot\delta{\bf A}_{1}({\bf r},t) \right)\;
\delta^{3}_{\rm gc} \]
follows from the definition \eqref{eq:psi_def}.

By substituting the variation \eqref{eq:delta_Hgy_phiA} in the variational principle derived from the gyrokinetic functional \eqref{eq:gy_action}, we obtain the gyrokinetic Maxwell's equations \cite{footnote}, which describe the self-consistent coupling between the gyrocenter Vlasov distribution $F$ and the electromagnetic perturbation fields ${\bf E}_{1}$ and ${\bf B}_{1}$:
\begin{eqnarray}
\epsilon\;\nabla\bdot{\bf E}_{1} & \equiv & 4\pi\,\int_{Z}\;F\;\fd{H_{\rm gy}}{\phi_{1}({\bf r})} \;=\; 4\pi\,\int_{Z}\,e\; F\;\left\langle {\sf T}_{\epsilon}^{-1}\,\delta^{3}_{\rm gc}\right\rangle, 
\label{eq:gy_Poisson} \\
\nabla\btimes({\bf B}_{0} + \epsilon\,{\bf B}_{1}) \;-\;\frac{\epsilon}{c}\;\pd{{\bf E}_{1}}{t} & \equiv & -\;4\pi\,\int_{Z}\;F\;
\fd{H_{\rm gy}}{{\bf A}_{1}({\bf r})} \;=\; 4\pi\,\int_{Z}\,e\; F\;\left\langle {\sf T}_{\epsilon}^{-1}\left(\frac{{\bf v}_{0}}{c}\;
\delta^{3}_{\rm gc}\right) \right\rangle,
\label{eq:gy_Ampere}
\end{eqnarray}
where we use the gyrocenter push-forward definitions
\begin{equation}
{\sf T}_{\epsilon}^{-1}\delta_{\rm gc}^{3} \;\equiv\; \delta_{\rm gc}^{3} \;+\; \epsilon\;\vb{\rho}_{1{\rm gy}}\bdot\nabla\delta_{\rm gc}^{3},
\label{eq:push_Poisson}
\end{equation}
and
\begin{equation}
{\sf T}_{\epsilon}^{-1}\left({\bf v}_{0}\frac{}{}\delta_{\rm gc}^{3}\right) \;\equiv\; \left( {\bf v}_{0} \;+\; \epsilon\;
\dot{{\bf R}}_{1{\rm gy}}\right) \;\delta_{\rm gc}^{3} \;+\; \epsilon \left( \delta_{\rm gc}^{3}\;\frac{d_{\rm gc}}{dt}\vb{\rho}_{1{\rm gy}} \;+\; 
{\bf v}_{0}\;\vb{\rho}_{1{\rm gy}}\bdot\nabla\delta_{\rm gc}^{3} \right).
\label{eq:push_Ampere}
\end{equation}
When the polarization and magnetization terms (associated with $\vb{\rho}_{1{\rm gy}}$) are separated from the gyrocenter source terms 
\cite{Brizard_Vlasovia}, the gyrokinetic Maxwell's equations \eqref{eq:gy_Poisson}-\eqref{eq:gy_Ampere} can also be expressed as
\begin{eqnarray}
\epsilon\;\nabla\bdot{\bf D}_{1} & \equiv & 4\pi\,\int_{Z}\,e\; F\;\left\langle \delta^{3}_{\rm gc}\right\rangle, 
\label{eq:gy_Poisson_D} \\
\nabla\btimes({\bf B}_{0} + \epsilon\,{\bf H}_{1}) \;-\;\frac{\epsilon}{c}\;\pd{{\bf D}_{1}}{t} & \equiv & 4\pi\,\int_{Z}\,e\; 
F\;\left( \left\langle \frac{{\bf v}_{0}}{c}\;\delta^{3}_{\rm gc} \right\rangle \;+\; \epsilon\;\dot{{\bf R}}_{1{\rm gy}}\;
\langle\delta_{\rm gc}^{3}\rangle \right),
\label{eq:gy_Ampere_HD}
\end{eqnarray}
where the reduced electromagnetic fields ${\bf D}_{1} \equiv {\bf E}_{1} + 4\pi\,{\bf P}_{1{\rm gy}}$ and ${\bf H}_{1} \equiv {\bf B}_{1} - 4\pi\,
{\bf M}_{1{\rm gy}}$ are defined in terms of the first-order gyrocenter polarization and magnetization (using $\delta_{\rm gc} \simeq \delta^{3}$)
\begin{equation}
\left( \begin{array}{c}
{\bf P}_{1{\rm gy}} \\
\\
{\bf M}_{1{\rm gy}}
\end{array} \right) \;\simeq\; \int\; F\; \left( \begin{array}{c}
\vb{\pi}_{1{\rm gy}} \\
\\
\vb{\mu}_{1{\rm gy}}
\end{array} \right) d^{3}p,
\label{eq:PM_gy_def}
\end{equation}
which are expressed in terms of gyrocenter phase-space moments of the first-order gyrocenter electric-dipole and magnetic-dipole moments 
\eqref{eq:pigy_def}-\eqref{eq:mugy_def}. We note that the gyrocenter source terms in Eqs.~\eqref{eq:gy_Poisson_D}-\eqref{eq:gy_Ampere_HD} now only include the effects due to guiding-center polarization $(\langle\delta_{\rm gc}^{3}\rangle)$ and magnetization $(\langle{\bf v}_{0}\,
\delta_{\rm gc}^{3}\rangle)$, in addition to the perturbed gyrocenter current density associated with the first-order gyrocenter velocity 
$\dot{{\bf R}}_{1{\rm gy}}$ defined in Eq.~\eqref{eq:R1dot}.

\subsection{Gyrokinetic Energy Conservation}

The exact \textit{global} gyrokinetic energy conservation law $dE_{\rm gy}/dt \equiv 0$ for the full-$f$ gyrokinetic Vlasov-Maxwell equations 
\eqref{eq:gy_Vlasov} and \eqref{eq:gy_Poisson}-\eqref{eq:gy_Ampere} is expressed in terms of the total gyrokinetic energy \cite{Brizard_89,Brizard_EPB}
\begin{equation}
E_{\rm gy} \;=\; \int \frac{d^{3}x}{8\pi} \left( \epsilon^{2}\;|{\bf E}_{1}|^{2} \;+\; |{\bf B}_{0} + \epsilon\,{\bf B}_{1}|^{2} \right) \;+\; \int_{Z} 
\;F\, \left( H_{\rm gy} \;-\; \epsilon\;e \left\langle {\sf T}_{\epsilon}^{-1}\phi_{1{\rm gc}} \right\rangle \right),
\label{eq:Egyro_def}
\end{equation}
which includes higher-order terms associated with the electric-field energy. This expression was derived by Noether method 
\cite{Brizard_VP1,Brizard_VP2} and, hence, one can explicitly show that $dE_{\rm gy}/dt = 0$ at all orders in $\epsilon$. The explicit proof presented heer emphasizes the role played by the second-order ponderomotive gyrocenter Hamiltonian, on the one hand, and the gyrocenter polarization and magnetization effects in the gyrokinetic Maxwell's equations, on the other hand.

\subsubsection{Gyrocenter kinetic energy}

The Vlasov term in Eq.~(\ref{eq:Egyro_def}) introduces the gyrocenter ``kinetic'' energy
\begin{eqnarray}
K_{\rm gy} & \equiv & \left\langle {\sf T}_{\rm gy}^{-1}H_{0{\rm gc}}\right\rangle \;=\; H_{\rm gy} \;-\; \epsilon\;e \left\langle 
{\sf T}_{\epsilon}^{-1}\phi_{1{\rm gc}} \right\rangle \nonumber \\
 & = & H_{0{\rm gc}} \;-\; \epsilon\,e\;\left\langle {\bf A}_{1{\rm gc}}\bdot\frac{{\bf v}_{0}}{c}\right\rangle \;+\; \epsilon^{2}\;e\,
\left\langle \pounds_{\rm gy}\left( \phi_{1{\rm gc}} \;-\; \frac{1}{2}\,\psi_{1{\rm gc}} \right) \right\rangle,
\label{eq:Kgy_def}
\end{eqnarray}
which removes the first-order contribution from $\phi_{1{\rm gc}}$ and alters the second-order ponderomotive Hamiltonian. By rearranging the second-order terms in Eq.~(\ref{eq:Kgy_def}), we obtain the convenient expression
\begin{equation}
K_{\rm gy} \;\equiv\; \frac{1}{2m}\;\left\langle\left|{\bf p}_{0} \;-\; \epsilon\;\frac{e}{c}\,{\bf A}_{1{\rm gc}}\right|^{2} \right\rangle \;+\;
\frac{\epsilon^{2}\,e^{2}}{2\,\Omega} \left\langle \left\{ \wt{\Phi}_{1{\rm gc}},\; \wt{\phi}_{1{\rm gc}} \right\}_{\rm gc} \;-\; \left\{ 
\wt{\Xi}_{1{\rm gc}},\; \wt{\xi}_{1{\rm gc}} \right\}_{\rm gc} \right\rangle,
\label{eq:Kgy_explicit}
\end{equation}
where we use the notation $(\wt{\Phi}_{1{\rm gc}},\,\wt{\Xi}_{1{\rm gc}}) \equiv \int\,(\wt{\phi}_{1{\rm gc}},\,\wt{\xi}_{1{\rm gc}})\,d\theta$, with 
$\xi_{1{\rm gc}} \equiv {\bf A}_{1{\rm gc}}\bdot{\bf v}_{0}/c$, and we introduced the gyrocenter ``canonical'' kinetic energy
\begin{equation} 
\frac{1}{2m}\;\left\langle\left|{\bf p}_{0} \;-\; \epsilon\;\frac{e}{c}\,{\bf A}_{1{\rm gc}}\right|^{2} \right\rangle \;\equiv\; \mu\,B_{0} \;+\;
\frac{p_{\|}^{2}}{2m} \;-\; \epsilon\;e\,\left\langle {\bf A}_{1{\rm gc}}\bdot\frac{{\bf v}_{0}}{c}\right\rangle \;+\; 
\frac{\epsilon^{2}\,e^{2}}{2\,mc^{2}}\;\left\langle|{\bf A}_{1{\rm gc}}|^{2}\right\rangle,
\label{eq:can_kinetic}
\end{equation}
defined in terms of the guiding-center momentum ${\bf p}_{0} \equiv m\,{\bf v}_{0}$, which satisfies the guiding-center identity \eqref{eq:vgc_2}. 

The last terms in Eq.~(\ref{eq:Kgy_explicit}) involve separate ponderomotive electric and magnetic contributions whose physical significance is explained as follows. First, we return to the expression \eqref{eq:Kgy_def} for the gyrocenter kinetic energy and, using the low-frequency definition
$\langle\pounds_{\rm gy}\phi_{1{\rm gc}}\rangle \simeq \langle\vb{\rho}_{1{\rm gc}}\bdot{\bf E}_{1{\rm gc}}\rangle$, we obtain another form of the gyrocenter kinetic energy \eqref{eq:Kgy_def}:
\begin{eqnarray}
K_{\rm gy} & = & H_{0{\rm gc}} \;-\; \epsilon\,\frac{e}{c} \left\langle {\bf A}_{1{\rm gc}}\bdot \left\{ {\bf R} + \vb{\rho}_{0},\; \left(
H_{0{\rm gc}} \;+\; \epsilon\,\frac{e}{2}\,\langle\psi_{1{\rm gc}}\rangle \right) \right\}_{\rm gc} \right\rangle \nonumber \\
 &  &+\; \frac{\epsilon^{2}}{2} \left\langle \left(e\frac{}{}\vb{\rho}_{1{\rm gy}}\right)\bdot{\bf E}_{1{\rm gc}} \;-\; \left(e\,\vb{\rho}_{1{\rm gy}}\btimes\frac{{\bf v}_{0}}{c}\right)\bdot{\bf B}_{1{\rm gc}} \right\rangle.
\label{eq:Kgy_polmag}
\end{eqnarray}
Here, we simply note that the negative sign between the polarization and magnetization energies is needed in order to recover the standard expression for the second-order macroscopic electromagnetic energy \cite{Jackson}
\begin{equation}
\frac{\epsilon^{2}}{8\pi} \left( {\bf E}_{1}\bdot{\bf D}_{1} \;+\frac{}{} {\bf B}_{1}\bdot{\bf H}_{1} \right) \;\equiv\; \frac{\epsilon^{2}}{8\pi} 
\left( |{\bf E}_{1}|^{2} \;+\; |{\bf B}_{1}|^{2} \right) \;+\; \frac{\epsilon^{2}}{2} \left( {\bf P}_{1{\rm gy}}\bdot{\bf E}_{1} \;-\; 
{\bf M}_{1{\rm gy}}\bdot{\bf B}_{1} \right),
\label{eq:DH_energy}
\end{equation}
where the first-order gyrocenter polarization and magnetization are defined in Eq.~\eqref{eq:PM_gy_def}.

\subsubsection{Proof of exact energy conservation}

We now proceed with the explicit proof of energy conservation $dE_{\rm gy}/dt \equiv 0$. In contrast to the previous proof of energy conservation 
\cite{Brizard_EPB} (which ignored the higher-order terms associated with ${\bf E}_{1} + c\,\nabla\phi_{1} \neq 0$), the proof presented below focuses on the role played by quadratic nonlinearities in the gyrocenter Hamiltonian \eqref{eq:gy_Hamiltonian}, here expressed as
\begin{equation}
H_{\rm gy} \;\equiv\; H_{0{\rm gc}} \;+\; \epsilon\,e\; \left( \langle \psi_{1{\rm gc}}\rangle \;-\; \frac{\wh{\epsilon}}{2}\;
\left\langle \pounds_{\rm gy}\,\psi_{1{\rm gc}}\right\rangle \right), 
\label{eq:H_epsdel}
\end{equation}
where we introduce the {\it nonlinearity} ordering parameter $\wh{\epsilon}$ in Eq.~(\ref{eq:H_epsdel}), which is either $\wh{\epsilon} = \epsilon$ (for a nonlinear gyrocenter Hamiltonian) or $\wh{\epsilon} = 0$ (for a truncated gyrocenter Hamiltonian). 

By defining the push-forward operator ${\sf T}_{\wh{\epsilon}}^{-1} \equiv 1 - \wh{\epsilon}\,\pounds_{\rm gy}$ associated with the nonlinearity parameter $\wh{\epsilon}$, the expression for $dE_{\rm gy}/dt$ is separated into four energy-exchange terms:
\begin{equation}
\frac{dE_{\rm gy}}{dt} \;\equiv\; {\cal P}_{F} \;+\; {\cal P}_{H} \;+\; {\cal P}_{\phi} \;+\; {\cal P}_{A},
\label{eq:Edot_gyro}
\end{equation}
where the Vlasov term is
\begin{equation}
{\cal P}_{F} \;\equiv\; \int_{Z}\;\pd{F}{t}\;H_{\rm gy}, 
\label{eq:E_I}
\end{equation}
the Hamiltonian term is
\begin{equation}
{\cal P}_{H} \;\equiv\; \int_{Z}\; F \left( \pd{H_{\rm gy}}{t} \;-\; \epsilon\,e\,\left\langle \pd{}{t}\left({\sf T}_{\wh{\epsilon}}^{-1}\psi_{1{\rm gc}}\right)\right\rangle \right), 
\label{eq:E_II}
\end{equation}
the Poisson term is
\begin{equation}
{\cal P}_{\phi} \;\equiv\; \int d^{3}x\,\left(\frac{\epsilon^{2}\phi_{1}}{4\pi}\right)\;\nabla\bdot\pd{{\bf E}_{1}}{t} \;-\; \epsilon \int_{Z}\,e\;
\pd{F}{t}\;\left\langle{\sf T}_{\wh{\epsilon}}^{-1}\phi_{1{\rm gc}}\right\rangle, 
\label{eq:E_III}
\end{equation}
and the Amp\`{e}re term is
\begin{equation}
{\cal P}_{A} \;\equiv\; \int d^{3}x \left(\frac{\epsilon}{4\pi}\;\pd{{\bf A}_{1}}{t}\right)\bdot\left( \nabla\btimes{\bf B} \;-\; \frac{\epsilon}{c}\;
\pd{{\bf E}_{1}}{t} \right) \;-\; \epsilon\;\int_{Z}\,e\; F\;\left\langle \pd{}{t}\left[\,{\sf T}_{\wh{\epsilon}}^{-1}\left({\bf A}_{1{\rm gc}}\bdot
\frac{{\bf v}_{0}}{c}\right)\,\right]\right\rangle. 
\label{eq:E_IV}
\end{equation}
Note that we obtain these energy-exchange terms directly from Eq.~(\ref{eq:Egyro_def}) by rearranging terms and integrating by parts (where surface terms are expected to vanish based on suitable boundary conditions). The separations shown in Eqs.~(\ref{eq:E_I})-(\ref{eq:E_IV}) combine terms of similar nature and facilitates the proof and its interpretation. The proof that $dE_{\rm gy}/dt \equiv 0$ in Eq.~\eqref{eq:Edot_gyro} uses the operator-commutation identity
\begin{equation}
\left[ {\sf T}_{\epsilon}^{-1},\; \pd{}{t}\right]\,g \;\equiv\; {\sf T}_{\epsilon}^{-1}\left(\pd{g}{t}\right) \;-\; \pd{}{t}\left(
{\sf T}_{\epsilon}^{-1}\frac{}{}g\right) \;=\; \epsilon\; \left( \left\{ \pd{S_{1}}{t},\; g \right\}_{\rm gc} \;+\; \frac{e}{c}\,
\pd{{\bf A}_{1{\rm gc}}}{t}\bdot\left\{ {\bf R} + \vb{\rho}_{0},\frac{}{} g \right\}_{\rm gc} \right)
\label{eq:time_T}
\end{equation}
between the partial time derivative operator and the push-forward operator (acting on an arbitrary function $g$), which involves the explicit time dependence of the gyrocenter transformation.

First, using the gyrocenter Vlasov equation (\ref{eq:gy_Vlasov}), the Vlasov term (\ref{eq:E_I}) can be written as
\begin{equation}
{\cal P}_{F} \;=\; -\;\int_{Z}\; \{(F\,H_{\rm gy}),\; H_{\rm gy}\}_{\rm gc} \;\equiv\; 0,
\label{eq:P_I_final}
\end{equation}
when we use general property $\int_{Z}\,\{ f, g\}_{\rm gc} \equiv 0$ of the guiding-center Poisson bracket (for arbitrary functions $f$ and $g$). We note that Eq.~(\ref{eq:P_I_final}) vanishes under all versions of the nonlinear gyrokinetic equations (i.e., for arbitrary $\wh{\epsilon}$). Second, we express the partial time derivative of the gyrocenter Hamiltonian (\ref{eq:gy_Hamiltonian}) in the Hamiltonian term 
(\ref{eq:E_II}), using Eq.~\eqref{eq:delta_Hgy}, as
\[ \pd{H_{\rm gy}}{t} \;=\; \epsilon\,e\; \left( \pd{\langle\psi_{1{\rm gc}}\rangle}{t} \;-\; \frac{\wh{\epsilon}}{2}\,\;\pd{}{t}\left\langle 
\pounds_{\rm gy}\,\psi_{1{\rm gc}}\right\rangle \right) \;\equiv\; \epsilon\,e\;\left\langle {\sf T}_{\wh{\epsilon}}^{-1}\;\left( \pd{\psi_{1{\rm gc}}}{t}\right)\right\rangle,  \]
and the Hamiltonian term (\ref{eq:E_II}) thus becomes
\begin{equation}
{\cal P}_{H} \;=\; \epsilon\,\int_{Z}\,e\; F \left[\; \left\langle {\sf T}_{\wh{\epsilon}}^{-1}\;\left( \pd{\psi_{1{\rm gc}}}{t}\right)\right\rangle \;-\; \left\langle \pd{}{t}\left({\sf T}_{\wh{\epsilon}}^{-1}\psi_{1{\rm gc}}\right)\right\rangle \;\right] \;\equiv\; \epsilon\;\int_{Z}\,e\; F 
\left\langle \left[ {\sf T}_{\wh{\epsilon}}^{-1},\; \pd{}{t}\right]\psi_{1{\rm gc}}\right\rangle,
\label{eq:P_II_final}
\end{equation}
where the operator-commutation definition (\ref{eq:time_T}) is used. Here, we note that this energy-exchange term is non-vanishing only when the gyrocenter Hamiltonian (\ref{eq:gy_Hamiltonian}) retains the second-order ponderomotive contribution $(\wh{\epsilon} \neq 0)$, i.e., $[{\sf T}_{0}^{-1},
\;\partial/\partial t] \equiv 0$. Third, we use the partial time derivative of the gyrokinetic Poisson's equation (\ref{eq:gy_Poisson}) to obtain
\[ \int d^{3}x\,\frac{\epsilon^{2}\phi_{1}}{4\pi}\;\nabla\bdot\pd{{\bf E}_{1}}{t} \;=\; \epsilon\;\int_{Z}\,e\; \left( \pd{F}{t}\; \left\langle
{\sf T}_{\epsilon}^{-1}\phi_{1{\rm gc}} \right\rangle \;-\; F\;\left\langle \left[{\sf T}_{\epsilon}^{-1},\;\pd{}{t}\right]\phi_{1{\rm gc}} 
\right\rangle \right), \]
so that the Poisson term (\ref{eq:E_III}) becomes
\begin{equation}
{\cal P}_{\phi} \;=\; -\; \epsilon\,\int_{Z}\,e\; F \left\langle \left[{\sf T}_{\epsilon}^{-1},\;\pd{}{t}\right]\phi_{1{\rm gc}} \right\rangle \;+\; \epsilon\,\int_{Z}\,e\;\pd{F}{t} \left( \left\langle{\sf T}_{\epsilon}^{-1}\phi_{1{\rm gc}} \right\rangle \;-\frac{}{} \left\langle
{\sf T}_{\wh{\epsilon}}^{-1}\phi_{1{\rm gc}} \right\rangle \right).
\label{eq:P_III_final} 
\end{equation}
Here, the first term on the right side of Eq.~\eqref{eq:P_III_final} involves the gyrocenter polarization in the gyrokinetic Poisson's equation
(\ref{eq:gy_Poisson}), while the remaining terms vanish for the nonlinear case $\wh{\epsilon} = \epsilon$. Fourth, we use the gyrokinetic Amp\`{e}re's equation (\ref{eq:gy_Ampere}) to obtain
\[ \int \frac{d^{3}x}{4\pi}\;\pd{{\bf A}_{1}}{t}\bdot\left( \nabla\btimes{\bf B} \;-\; \frac{\epsilon}{c}\;\pd{{\bf E}_{1}}{t} \right) \;=\; 
\int_{Z}\,e\; F \left\langle{\sf T}_{\epsilon}^{-1}\left(\pd{{\bf A}_{1{\rm gc}}}{t}\bdot\frac{{\bf v}_{0}}{c}\right)\right\rangle, \]
and we use the operator-commutation formula (\ref{eq:time_T}), with $g \equiv \xi_{1{\rm gc}}$, to obtain the Amp\`{e}re term
\begin{widetext}
\begin{equation} 
{\cal P}_{A} \;=\; \epsilon\int_{Z}\,e\; F \left\langle \left[\; {\sf T}_{\epsilon}^{-1},\; \pd{}{t} \right] \xi_{1{\rm gc}} \right\rangle \;+\; \epsilon\,\int_{Z}\,e\; F \left[ \left\langle \pd{}{t}\left({\sf T}_{\epsilon}^{-1}\frac{}{}\xi_{1{\rm gc}} \right) \right\rangle - \left\langle 
\pd{}{t}\left({\sf T}_{\wh{\epsilon}}^{-1}\xi_{1{\rm gc}}\right) \right\rangle \;\right].
\label{eq:P_IV_final}
\end{equation}
Here, the first term on the right side of Eq.~\eqref{eq:P_IV_final} involves the gyrocenter magnetization in the gyrokinetic Amp\`{e}re's equation 
(\ref{eq:gy_Ampere}), while the remaining terms vanish for the nonlinear case $\wh{\epsilon} = \epsilon$. 

By combining Eqs.~(\ref{eq:P_I_final})-(\ref{eq:P_IV_final}) into Eq.~(\ref{eq:Edot_gyro}), we obtain
\begin{equation}
\frac{dE_{\rm gy}}{dt} \;\equiv\; \epsilon\,(\wh{\epsilon} - \epsilon)\;\int_{Z}\,e\; \left[\; \pd{F}{t}\;\left\langle \pounds_{\rm gy}\,\phi_{1{\rm gc}}\right\rangle 
\;+\; F \left\langle \pd{}{t}\;\left(\pounds_{\rm gy}\frac{}{}\xi_{1{\rm gc}}\right)\right\rangle \;-\; F\;\left\langle \left[ 
\pounds_{\rm gy},\;\pd{}{t}\right]\psi_{1{\rm gc}}\right\rangle \;\right],
\label{eq:Edot_final}
\end{equation}
\end{widetext}
where we expanded the push-forward operator ${\sf T}_{\epsilon}^{-1} = 1 - \;\epsilon\,\pounds_{\rm gy}$ (associated with the gyrokinetic polarization and magnetization) and the push-forward operator ${\sf T}_{\wh{\epsilon}}^{-1} = 1 - \;\wh{\epsilon}\,\pounds_{\rm gy}$ (associated with the nonlinear gyrocenter Hamiltonian). Equation (\ref{eq:Edot_final}) clearly shows that only the case of a nonlinear gyrocenter Hamiltonian ($\wh{\epsilon} = 
\epsilon$) satisfies the exact gyrokinetic energy conservation law $dE_{\rm gy}/dt = 0$. The gyrokinetic energy conservation law also implies that the energy-transfer processes can be represented in terms of the pair-wise couplings
\begin{equation}
{\cal P}_{\phi} \;\leftrightarrow\; {\cal P}_{H} \;\leftrightarrow\; {\cal P}_{A},
\label{eq:phi_H_A}
\end{equation}
where electric and magnetic fluctuations separately contribute to the evolution of the gyrocenter Hamiltonian (and thus to the evolution of the gyrocenter Vlasov distribution). 

Lastly, we note that if the nonlinearity parameter $\wh{\epsilon}$ is set to zero in the gyrocenter Hamiltonian 
\eqref{eq:H_epsdel}, we obtain $dE_{\rm gy}/dt = {\cal O}(\epsilon^{2})$, i.e., the non-conservation of gyrokinetic energy arises at the same order as the nonlinear time scale associated with the turbulent gyrokinetic dynamics, which can lead to unphysical saturation levels \cite{Brizard_Hahm}. In particular, the Hamiltonian energy transfer term \eqref{eq:P_II_final} vanishes for $\wh{\epsilon} = 0$, which interferes with the nonlinear energy-transfer processes \eqref{eq:phi_H_A}.

\section{\label{sec:trunc}Truncated Hamiltonian Gyrokinetic Formulation}

Most numerical simulations of the nonlinear gyrokinetic Vlasov-Poisson and Vlasov-Maxwell equations ignore the quadratic gyrocenter Hamiltonian 
in the gyrokinetic Vlasov equation [i.e., $\wh{\epsilon} \equiv 0$ in Eq.~(\ref{eq:H_epsdel})]. Under these conditions, it is important to ask whether there still exists an exact gyrokinetic energy conservation law. The answer is yes but only if we use an explicit truncation scheme in the variational principle (\ref{eq:gy_action}) for the nonlinear gyrokinetic equations. 

\subsection{Truncated gyrokinetic Vlasov-Maxwell equations}

In the truncated ($\delta f$) version, the gyrocenter Vlasov distribution is expressed as $F \equiv F_{0} + \epsilon\,F_{1}$, where $F_{0}$ is the reference distribution function and $F_{1}$ denotes the departure from $F_{0}$ driven by linear and nonlinear effects. Here, $F_{0}$ is assumed to be time independent and satisfies the equilibrium condition $\{ F_{0},\;H_{0}\}_{\rm gc} \equiv 0$, i.e., it is an arbitrary function, not necessarily Maxwellian, of the constants of the motion for guiding-center Hamiltonian dynamics.

The truncated gyrokinetic (trgy) Vlasov equation is written as
\begin{equation} 
0 \;\equiv\; \epsilon\;\frac{d_{\rm gc}F_{1}}{dt} \;+\; \epsilon \left\{ \left(F_{0} \;+\frac{}{} 
\epsilon\;F_{1}\right),\; e\;\langle\psi_{1{\rm gc}}\rangle \right\}_{\rm gc},
\label{eq:gy_Vlasov_delta}
\end{equation}
where $d_{\rm gc}/dt \equiv \partial/\partial t + \{\;,\; H_{0{\rm gc}}\}_{\rm gc}$ denotes the (unperturbed) guiding-center Vlasov operator, the 
linear drive is $\epsilon\,\{ F_{0},\; e\,\langle\psi_{1{\rm gc}}\rangle\}_{\rm gc}$ and the nonlinear drive 
\begin{equation}
\{ F_{1},\; e\,\langle\psi_{1{\rm gc}}\rangle\}_{\rm gc} \;\equiv\; \frac{c\bhat_{0}}{B_{0}}\btimes\nabla\langle\psi_{1{\rm gc}}\rangle\bdot\nabla F_{1}
\;+\; e\;\bhat_{0}\bdot\left(\nabla F_{1}\;\pd{\langle\psi_{1{\rm gc}}\rangle}{p_{\|}} \;-\; \pd{F_{1}}{p_{\|}}\;\nabla\langle\psi_{1{\rm gc}}\rangle
\right), 
\label{eq:truncated_nonlinear}
\end{equation}
which combines the perpendicular nonlinear drive and the so-called parallel nonlinearity. The perturbed gyrocenter Vlasov distribution $F_{1}$ is coupled to the truncated gyrokinetic Maxwell equations
\begin{eqnarray}
\epsilon\;\nabla^{2}\phi_{1} & = & -\,4\pi\,\int_{Z}\,e\; \left[\; (F_{0} + \epsilon\;F_{1})\;\left\langle\delta^{3}_{\rm gc}\right\rangle \;-\frac{}{} \epsilon\;F_{0}\;\left\langle \pounds_{\rm gy}\delta^{3}_{\rm gc}\right\rangle \;\right], 
\label{eq:gy_Poisson_delta} \\
\nabla\btimes\left({\bf B}_{0} + \epsilon\,{\bf B}_{1}\right) & = & 4\pi\,\int_{Z}\,\frac{e}{c}\; \left[\; \left(F_{0} + \epsilon\,F_{1}\right)\;\left\langle {\bf v}_{0}\frac{}{}\delta^{3}_{\rm gc}\right\rangle \;-\; \epsilon\;F_{0}\;\left\langle \pounds_{\rm gy}\left({\bf v}_{0}
\frac{}{}\delta^{3}_{\rm gc}\right)\right\rangle \;\right],
\label{eq:gy_Ampere_delta}
\end{eqnarray}
where the effects of gyrocenter polarization and magnetization are represented by the gyrocenter operator $\pounds_{\rm gy}$. Here, the terms 
$\langle\delta^{3}_{\rm gc}\rangle$ and $\langle{\bf v}_{0}\delta^{3}_{\rm gc}\rangle$ are associated with {\it guiding-center} polarization and magnetization, respectively, while $\langle\pounds_{\rm gy}\delta^{3}_{\rm gc}\rangle$ and $\langle\pounds_{\rm gy}({\bf v}_{0}\delta^{3}_{\rm gc})\rangle$ are associated with {\it gyrocenter} polarization and magnetization.

The truncated gyrokinetic Vlasov-Maxwell equations (\ref{eq:gy_Vlasov_delta})-(\ref{eq:gy_Ampere_delta}) are obtained from a truncated variational principle based on the action functional
\begin{eqnarray}
{\cal A}_{\rm trgy} & \equiv & \int\frac{d^{4}x}{8\pi}\;\left( \epsilon^{2}\,|\nabla\phi_{1}|^{2} \;-\; |{\bf B}_{0} + \epsilon\,{\bf B}_{1}|^{2} 
\right) \;+\; \frac{\epsilon^{2}}{2}\;\int_{Z}\;e\,F_{0}\;\left\langle \pounds_{\rm gy}\;\psi_{1{\rm gc}}\right\rangle \nonumber \\
 &  &-\; \int\; {\cal F}_{\rm trgy}\;\left(H_{0{\rm gc}} \;+\frac{}{} \epsilon\;e\,\langle\psi_{1{\rm gc}}\rangle \;-\; W\right) \;d^{8}Z,
\label{eq:tr_gy}
\end{eqnarray}
where the last term represents the truncated gyrocenter Vlasov part, with ${\cal F}_{\rm trgy} \equiv (F_{0} + \epsilon\,F_{1})\;
\delta(W - H_{\rm trgy})$ denoting the truncated gyrocenter Vlasov distribution in extended phase space (with $H_{\rm trgy} \equiv H_{0{\rm gc}} + \epsilon\,e\,\langle\psi_{1{\rm gc}}\rangle$). We note that the term ${\cal F}\,H_{2{\rm gy}}$ in the full gyrokinetic action functional 
(\ref{eq:gy_action}) is replaced with $F_{0}\,H_{2{\rm gy}}$ in Eq.~(\ref{eq:tr_gy}), and we have inserted the low-frequency approximation ${\bf E}_{1} \equiv -\,\nabla\phi_{1}$. The second integral term (involving only the background Vlasov distribution $F_{0}$), therefore, does not contribute to the variational derivation of the truncated gyrokinetic Vlasov equation (\ref{eq:gy_Vlasov_delta}). Instead, it only contributes to the gyrocenter polarization and magnetization effects in the truncated gyrokinetic Maxwell's equations 
\eqref{eq:gy_Poisson_delta}-\eqref{eq:gy_Ampere_delta}. 

\subsection{Exact energy conservation law for truncated gyrokinetic equations}

The exact gyrokinetic energy conservation law $dE_{\rm trgy}/dt \equiv 0$ for the truncated gyrokinetic Vlasov-Maxwell equations 
(\ref{eq:gy_Vlasov_delta})-(\ref{eq:gy_Ampere_delta}), which is derived by Noether method from the variational principle (\ref{eq:tr_gy}). The truncated total gyrokinetic energy is
\begin{equation}
E_{\rm trgy} \;\equiv\; \int_{Z}\; \left[\; \left( F_{0} + \epsilon\,F_{1}\right) \left( H_{\rm trgy} \;-\frac{}{} 
\epsilon\,e\;\langle\phi_{1{\rm gc}} \rangle \right) \;+\; \epsilon^{2}\;F_{0}\;K_{2{\rm gy}} \right] \;+\; \int \frac{d^{3}x}{8\pi} \left( \epsilon^{2}\;|\nabla\phi_{1}|^{2} \;+\frac{}{} |{\bf B}_{0} + \epsilon\,{\bf B}_{1}|^{2} \right).
\label{eq:Egyro_delta}
\end{equation}
where the second-order gyrocenter kinetic energy is
\begin{eqnarray}
K_{2{\rm gy}} & \equiv & H_{2{\rm gy}} \;+\; e\;\langle\pounds_{\rm gy}\phi_{1{\rm gc}}\rangle \nonumber \\
 & = & \frac{e^{2}}{2\,mc^{2}}\;\left\langle|{\bf A}_{1{\rm gc}}|^{2}\right\rangle \;+\; \frac{e^{2}}{2\,\Omega} \left( 
\left\langle \left\{ \wt{\Phi}_{1{\rm gc}},\; \wt{\phi}_{1{\rm gc}} \right\}_{\rm gc} \;-\; \left\{ \wt{\Xi}_{1{\rm gc}},\; \wt{\xi}_{1{\rm gc}} 
\right\}_{\rm gc} \right\rangle \right).
\label{eq:K2_def}
\end{eqnarray}

The expression for $dE_{\rm trgy}/dt$ is again separated into four energy-exchange terms:
\begin{equation}
\frac{dE_{\rm trgy}}{dt} \;\equiv\; {\cal P}_{{\rm tr}F} \;+\; {\cal P}_{{\rm tr}H} \;+\; {\cal P}_{{\rm tr}\phi} \;+\; {\cal P}_{{\rm tr}A},
\label{eq:Edot_delta}
\end{equation}
where 
\begin{eqnarray}
{\cal P}_{{\rm tr}F} & = & \epsilon\;\int_{Z}\;( H_{0{\rm gc}} + \epsilon\,e\langle\psi_{1{\rm gc}}\rangle)\;\pd{F_{1}}{t} \;\equiv\; 0, 
\label{eq:trgy_F} \\ 
{\cal P}_{{\rm tr}\phi} & = & \epsilon^{2}\,\int_{Z}\,e\; F_{0} \left\langle \left[\pounds_{\rm gy},\; \pd{}{t}\right] 
\phi_{1{\rm gc}}\right\rangle, \label{eq:trgy_phi} \\
{\cal P}_{{\rm tr}A} & = & -\,\epsilon^{2}\int_{Z}\,e\;F_{0} \left\langle \pounds_{\rm gy}\left(\pd{{\bf A}_{1{\rm gc}}}{t}\bdot
\frac{{\bf v}_{0}}{c}\right) \right\rangle, \label{eq:trgy_A} \\
{\cal P}_{{\rm tr}H} & = & \epsilon^{2}\;\int_{Z}\; F_{0}\;\pd{K_{2{\rm gy}}}{t} \;\equiv\; -\,{\cal P}_{{\rm tr}\phi} \;-\; {\cal P}_{{\rm tr}A}.
\label{eq:trgy_H}
\end{eqnarray}
Note that Eq.~\eqref{eq:trgy_phi} is obtained from Eq.~\eqref{eq:P_III_final} by keeping only the first term while Eq.~\eqref{eq:trgy_A} is obtained from Eq.~\eqref{eq:P_IV_final} by setting $\wh{\epsilon} = 0$ in the last term and partially cancelling the remaining terms. In Eq.~(\ref{eq:trgy_H}), we used the relation
\begin{eqnarray*} 
\pd{K_{2{\rm gy}}}{t} & = & \left( \pd{H_{2{\rm gy}}}{t} \;+\; e\;\left\langle\pounds_{\rm gy}\pd{\phi_{1{\rm gc}}}{t}\right\rangle \right) \;+\; e\;\left\langle \left\{ \pd{S_{1}}{t},\; \phi_{1{\rm gc}} \right\}_{\rm gc}\right\rangle \\
 & \equiv & -\;e\,\left\langle\pounds_{\rm gy}\left( \pd{\psi_{1{\rm gc}}}{t} \;-\; \pd{\phi_{1{\rm gc}}}{t} \right)\right\rangle \;-\; e\;
\left\langle\left[\pounds_{\rm gy},\; \pd{}{t}\right]\phi_{1{\rm gc}} \right\rangle,
\end{eqnarray*}
and Eqs.~(\ref{eq:trgy_phi})-(\ref{eq:trgy_A}) for ${\cal P}_{{\rm tr}\phi}$ and ${\cal P}_{{\rm tr}A}$. Hence, we obtain $dE_{\rm trgy}/dt \equiv 0$ in Eq.~\eqref{eq:Edot_delta} and it is thus possible to construct a truncated version of the nonlinear gyrokinetic Vlasov-Maxwell equations that conserves energy exactly. Note that the energy-transfer processes \eqref{eq:phi_H_A} are now replaced by the truncated nonlinear processes ${\cal P}_{{\rm tr}\phi} \leftrightarrow {\cal P}_{{\rm tr}H} \leftrightarrow {\cal P}_{{\rm tr}A}$ expressed in terms of the background distribution $F_{0}$.

\subsection{Limiting forms}

As a last topic, we now discuss useful limiting forms of the truncated total gyrokinetic energy (\ref{eq:Egyro_delta}). In particular, we discuss the use of a truncated form of the guiding-center Poisson bracket, which leads to simplified expressions for the gyrocenter kinetic energy 
\eqref{eq:Kgy_explicit}. 

In all expressions involving the guiding-center Poisson bracket $\{\;,\;\}_{\rm gc}$ appearing in the polarization and magnetization contributions in the gyrokinetic Maxwell equations (\ref{eq:gy_Poisson_delta})-(\ref{eq:gy_Ampere_delta}), it is common to use the approximation 
\[ \langle\{f,\; g\}_{\rm gc}\rangle \;\simeq\; \frac{\Omega}{B}\,\pd{}{\mu} \left\langle g\;\pd{f}{\theta}\right\rangle, \]
where $f$ and $g$ are arbitrary functions, and the Poisson-bracket terms associated with the slower parallel and drift dynamics are omitted. Under this approximation, the second-order gyrocenter kinetic energy (\ref{eq:K2_def}) becomes
\begin{equation}
K_{2{\rm gy}} \;\simeq\; \frac{e^{2}}{2mc^{2}}\;\left\langle|{\bf A}_{1{\rm gc}}|^{2}\right\rangle \;+\; \frac{e^{2}}{2B_{0}}\;\pd{}{\mu} \left[\; \left(
\left\langle \phi_{1{\rm gc}}^{2}\right\rangle - \left\langle \phi_{1{\rm gc}}\right\rangle^{2} \right) \;-\; \left( \left\langle \xi_{1{\rm gc}}^{2}\right\rangle - \left\langle \xi_{1{\rm gc}}\right\rangle^{2} \right) \;\right],
\label{eq:K2_EM}
\end{equation}
where we used the identity $\langle \wt{\chi}_{1{\rm gc}}^{2}\rangle \equiv \langle\chi_{1{\rm gc}}^{2}\rangle - \langle\chi_{1{\rm gc}}\rangle^{2}$ for $\chi_{1{\rm gc}} = \phi_{1{\rm gc}}$ and $\xi_{1{\rm gc}}$. The total gyrokinetic energy (\ref{eq:Egyro_delta}) therefore becomes
\begin{eqnarray}
E_{\rm trgy} & \equiv & \int_{Z}\; \left[\; \left( F_{0} + \epsilon\,F_{1}\right) \left( H_{0{\rm gc}} \;-\frac{}{} 
\epsilon\,e\;\langle\xi_{1{\rm gc}} \rangle \right) \;+\; \epsilon^{2}\;\frac{e^{2}\,F_{0}}{2\;mc^{2}}\;\left\langle|{\bf A}_{1{\rm gc}}|^{2}\right\rangle \right] \nonumber \\
 &  &+\; \epsilon^{2}\; \left[\; \int \frac{d^{3}x}{8\pi} |\nabla\phi_{1}|^{2} \;+\; \int_{Z}\;\frac{e^{2}\,F_{0}}{2\,T} \left( \left\langle 
\phi_{1{\rm gc}}^{2}\right\rangle - \left\langle \phi_{1{\rm gc}}\right\rangle^{2} \right) \;\right] \nonumber \\
 &  &+\; \left[\; \int \frac{d^{3}x}{8\pi}\;|{\bf B}_{0} + \epsilon\,{\bf B}_{1}|^{2} \;-\; \epsilon^{2}\; \int_{Z}\;\frac{e^{2}\,F_{0}}{2\,T} \left( \left\langle \xi_{1{\rm gc}}^{2}\right\rangle - \left\langle \xi_{1{\rm gc}}\right\rangle^{2} \right) \;\right],
\label{eq:Egyro_delta_Max}
\end{eqnarray}
where we have assumed that the background gyrocenter Vlasov distribution $F_{0}$ is Maxwellian in $\mu$ (with temperature $T$). Similar limiting forms have been discussed elsewhere by Hahm {\it et al.} \cite{HLB,HWM}. Here, we easily recognize three sets of terms, each set with an immediate physical interpretation. The first group of terms corresponds to the truncated form of the canonical kinetic energy \eqref{eq:can_kinetic}, the second group (involving $\phi_{1}$) includes the electric-field energy and its gyrocenter polarization correction, while the third group (involving ${\bf A}_{1}$) includes the magnetic-field energy and its gyrocenter magnetization correction.

\section{\label{sec:summ}Summary and Future Work}

In conclusion, we have presented explicit proofs of the exact energy conservation laws for the full-$f$ and truncated-$\delta f$ versions of nonlinear gyrokinetic Vlasov-Maxwell equations. These proofs relied on the intimate connection between the gyrocenter ponderomotive Hamiltonian and the gyrocenter polarization and magnetization effects. In Sec.~\ref{sec:gyro_polmag}, we introduced the first-order gyrocenter gyroradius \eqref{eq:rhogy_def} in terms of which the gyrocenter polarization and magnetization effects are introduced explicitly in the gyrocenter Hamiltonian \eqref{eq:Hgy_polmag} and the gyrokinetic Maxwell equations \eqref{eq:gy_Poisson_D}-\eqref{eq:gy_Ampere_HD}. 

In Secs.~\ref{sec:gyro_Ham} and \ref{sec:gyro_polmag}, we showed that each dynamical reduction associated with unperturbed (guiding-center) and perturbed (gyrocenter) gyromotion introduces polarization and magnetization effects into the reduced Hamiltonian and the reduced Maxwell's equations. This connection between dynamical reduction and polarization-magnetization effects has been studied in the general context \cite{Brizard_Vlasovia}, in the high-frequency oscillation-center reduction \cite{Brizard_JCPS}, and in the case of the bounce-motion reduction \cite{Wang_Hahm}. In Secs.~\ref{sec:full} and \ref{sec:trunc}, we elucidated the connection between exact energy conservation and the accurate treatment of gyrocenter polarization and magnetization.

Future work will look at the prospect of constructing a variational principle from which the energy-conserving generalization of the nonlinear Frieman-Chen equations \cite{FC} might be derived.

\acknowledgments

The Author acknowledges useful discussions with Drs.~B.~D.~Scott and T.~S.~Hahm, and Prof.~Liu Chen. He is also grateful to the Institut de Recherche sur la Fusion par Confinement Magn\'{e}tique at CEA Cadarache, where this work was completed, for the kind hospitality. This work was partially supported by 
a U.~S.~Department of Energy grant No.~DE-FG02-09ER55005 and partially supported by the European Communities under the contract of Association between EURATOM and CEA, which was carried out within the framework of the European Fusion Development Agreement. The views and opinions expressed herein do not necessarily reflect those of the European Commission.


\begin{thebibliography}{99}
\bibitem{moving_mag} We note that the moving magnetic-dipole contribution to the electric dipole moment represents a relativistic correction that is not considered here; see Ref.~\cite{Brizard_JCPS} for a variational derivation.
\end{thebibliography}
\end{document}